\RequirePackage[l2tabu,orthodox]{nag}
\documentclass[12pt]{article}

\usepackage{mathtools}
\usepackage{amsmath}
\allowdisplaybreaks
\numberwithin{equation}{section}
\usepackage{amssymb}
\usepackage{amsfonts}
\usepackage{mathrsfs}
\usepackage[mathscr]{euscript}
\usepackage{amstext} 
\usepackage[a4paper,margin=1.0in]{geometry} 
\usepackage{graphicx,graphics,subfig}
\usepackage{epstopdf}
\usepackage{microtype}
\usepackage{float}
\usepackage[colorlinks=true,breaklinks,citecolor=DarkGreen,linkcolor=DarkRed,urlcolor=DarkBlue,menucolor=Blue,anchorcolor=DarkRed,
filecolor=Black,pdfauthor=author,linktocpage,pdfpagelabels,pdfborder={0 0 0},hyperfootnotes=false,bookmarks=true,
bookmarksnumbered=true]{hyperref}
\usepackage{cleveref}
\usepackage{algorithm}
\usepackage{setspace} 
\usepackage{latexsym} 
\usepackage[T1]{fontenc} 
\usepackage[square,comma,numbers,sort&compress]{natbib}
\usepackage[utf8]{inputenc} 
\usepackage[english]{babel} 
\usepackage[margin=10pt,font={small,stretch=1.1},labelfont=bf]{caption} 
\usepackage[pdftex,usenames,dvipsnames,svgnames,x11names,table]{xcolor} 
\usepackage{physics}
\usepackage[titles,subfigure]{tocloft}
\hypersetup{linktoc=all}
\usepackage{setspace}
\begin{document}

\begin{titlepage}
\hypersetup{pageanchor=false}
\renewcommand{\thefootnote}{\textcolor{DarkGreen}{\fnsymbol{footnote}}}

\begin{center}

\vspace*{1cm}
\begin{spacing}{2.0}
{\LARGE \textbf{Cone Holography with Neumann Boundary Conditions and Brane-localized Gauge Fields}}
\end{spacing}

\vskip 5mm

\vskip 10mm
{\large Zheng-Quan Cui\footnote{Email: \texttt{cuizhq5@mail.sysu.edu.cn}},
\large Yu Guo\footnote{Email: \texttt{guoy225@mail2.sysu.edu.cn}}, and
\large Rong-Xin Miao\footnote{Email: \texttt{miaorx@mail.sysu.edu.cn}}}

\vskip 5mm
{
\textit{School of Physics and Astronomy, Sun Yat-Sen University,\\ Zhuhai 519082, China}\\
}

\vskip 10mm

\today

\end{center}

\vskip 10mm
\begin{center}
{\textbf{\textsc{Abstract}}}
\end{center}
Cone holography is a codimension-$n$ doubly holographic model, which can be interpreted as the holographic dual of edge modes on defects. The initial model of cone holography is based on mixed boundary conditions. This paper formulates cone holography with Neumann boundary conditions, where the brane-localized gauge fields play an essential role. Firstly, we illustrate the main ideas in an AdS$_4$/CFT$_1$ toy model. We show that the $U(1)$ gauge field on the end-of-the-world brane can make the typical solution consistent with Neumann boundary conditions. Then, we generalize the discussions to general codimension-$n$ cone holography by employing brane-localized $p$-form gauge fields. We also investigate perturbative solutions and prove the mass spectrum of Kaluza-Klein gravitons is non-negative. Furthermore, we prove that cone holography obeys holographic $c$-theorem. Finally, inspired by the recently proposed chiral model in AdS/BCFT, we construct another type of cone holography with Neumann boundary conditions by applying massive vector (Proca) fields on the end-of-the-world brane. 
\vskip 5mm

\end{titlepage}

\newpage
\hypersetup{pageanchor=true}
\pagenumbering{arabic}

\renewcommand*{\thefootnote}{\arabic{footnote}}
\setcounter{footnote}{0}

\begin{spacing}{1.2}
\hypersetup{linkcolor=Black,filecolor=DarkGreen,urlcolor=DarkBlue}
\tableofcontents
\end{spacing}

\section{Introduction}
\label{sec:intro}

The AdS/CFT correspondence~\cite{Maldacena:1999m,Gubser:1998gkp,Witten:1998w} provides a fruitful framework for understanding the nature of gravity and strongly coupled gauge theories, which has been used in many branches of physics, e.g., condensed matter physics, quantum chromodynamics, hydrodynamics, cosmology, black hole information paradox, etc. Since a real physical system generally has boundaries, a natural generalization of the AdS/CFT correspondence is considering the holographic dual of boundary conformal field theory (BCFT), namely the AdS/BCFT correspondence~\cite{Takayanagi:2011zk,Fujita:2011ftt,Nozaki:2012qd} \footnote{See also~\cite{Miao:2018qkc,Miao:2017gyt,Chu:2017aab,Chu:2021mvq} for AdS/BCFT with various new boundary conditions.}. AdS/BCFT proposes that a BCFT on the AdS boundary is dual to the gravity coupled with an end-of-the-world (EOW) brane in bulk. It is a powerful tool to explore boundary effects such as the Casimir effect~\cite{Miao:2017aba} and anomalous transports~\cite{Chu:2018ntx,Miao:2022oas,Zhao:2023rno}. It is closely related to the brane world holography~\cite{Randall:1999ee,Randall:1999vf,Karch:2000ct} and the so-called doubly holography, which plays a vital role in the recent breakthrough of black hole information paradox by the island prescription~\cite{Penington:2019npb,Almheiri:2019psf,Almheiri:2019hni}. The island surface terminates on an EOW brane and enables a novel phase transition in recovering the Page curve of Hawking radiations. In recent few years, the island prescription has been explored from many aspects~\cite{Almheiri:2019yqk,
Almheiri:2019psy,Chen:2020uac,Chen:2020hmv,Ling:2020laa,Geng:2020qvw,Geng:2023zhq,Krishnan:2020fer,Yadav:2022mnv,Emparan:2023dxm,Kawabata:2021hac,Chou:2021boq,Alishahiha:2020qza,Hu:2022ymx,Hu:2022zgy,Miao:2022mdx,Miao:2023unv,Li:2023fly,Jeong:2023hrb,Yu:2023whl,Chang:2023gkt,Tong:2023nvi,Ghodrati:2022hbb,Lee:2022efh,Aguilar-Gutierrez:2023tic,Guo:2023fly,Liu:2023ggg,Lin:2023ajt,Lin:2023hzs}.

\begin{figure}[htbp]
\centering
\subfloat[ \ Wedge holography (WH) \label{sfig:wedge}]{
\includegraphics[width=2.5in]{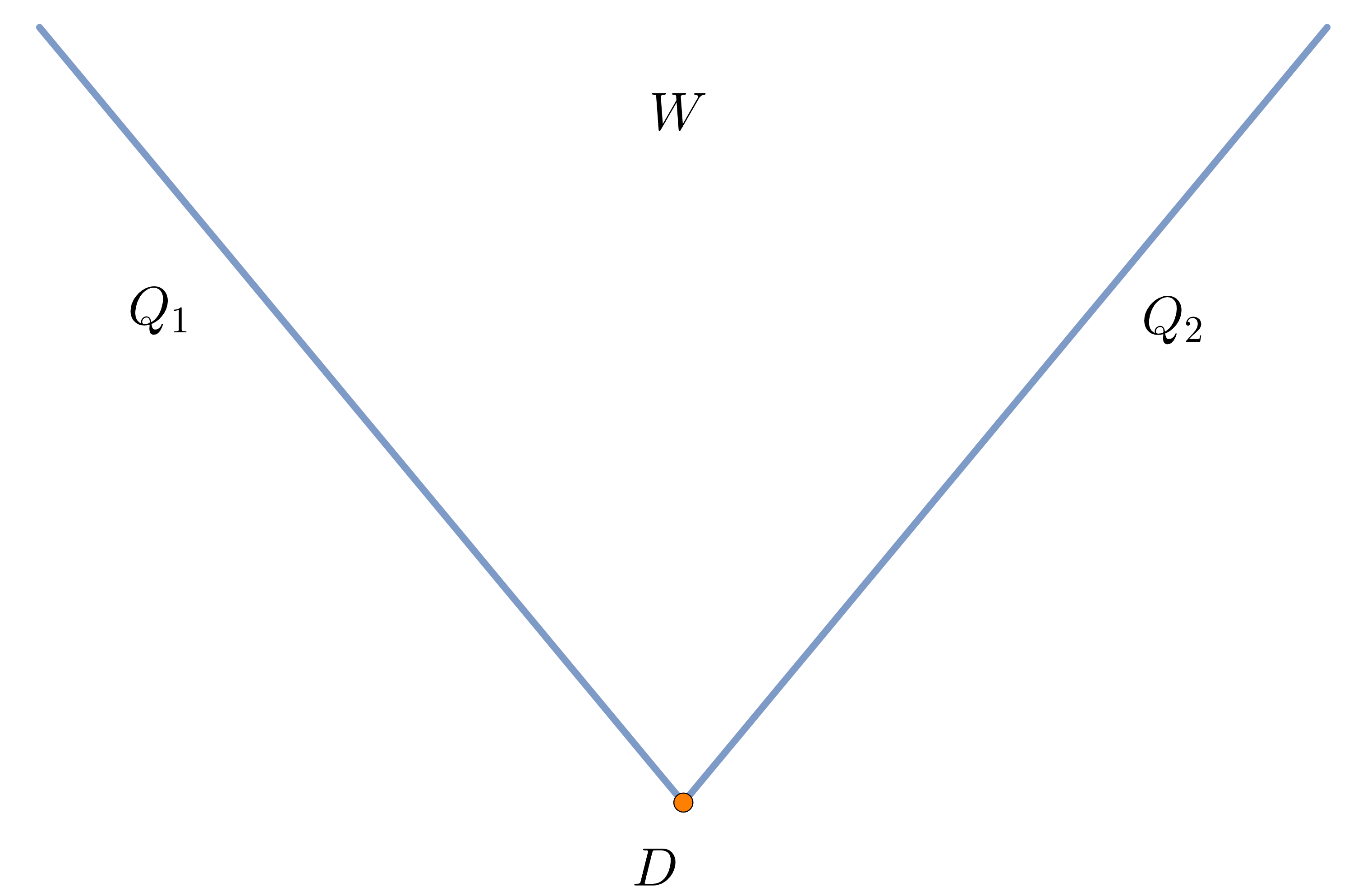}}
\hspace{0.5in}
\subfloat[\ WH from AdS/BCFT\label{sfig:adsbcft}]{
\includegraphics[width=2.5in]{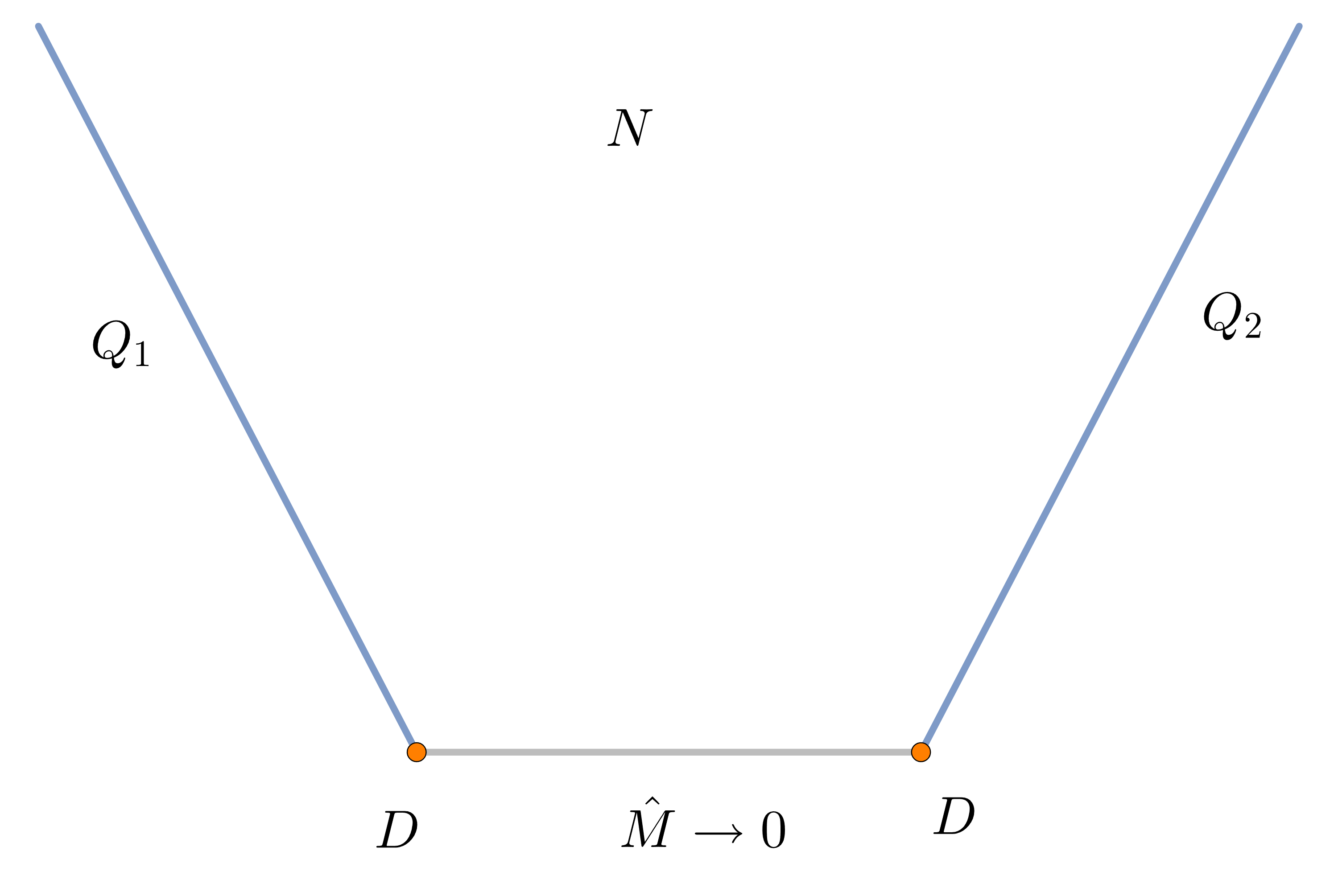}}
\caption{(left) Geometry of wedge holography; (right) Wedge holography from AdS/BCFT. In the left figure, $Q=Q_1\cup Q_2$ are EOW branes, $W$ is the bulk wedge space bound by $Q$, i.e., $\partial W=Q$, $D=\partial Q_1=\partial Q_2$ is the corner of wedge.  In the right figure, $N$ labels the bulk AdS space, $\hat{M}$ is the AdS boundary where BCFT lives, $D$ denotes the boundary (defect) of BCFT. In the limit $\hat{M}\to 0$, the bulk spacetime $N$ becomes a wedge $W$ and we obtain wedge holography from AdS/BCFT.}\label{wedgeholography}
\end{figure}

As a generalization of AdS/CFT and doubly holography, a novel codimension-$2$ holography called wedge holography was proposed in Ref.~\cite{Akal:2020wfl}. It is suggested that the classical gravity in a wedge bulk is dual to the quantum gravity on the EOW brane and is dual to a conformal field theory on the corner of the wedge. See Fig.~\ref{wedgeholography} (left) for the geometry. Wedge holography can be obtained as a special limit of AdS/BCFT; see Fig.~\ref{wedgeholography} (right). Wedge holography has passed non-trivial tests from entanglement/R\'enyi entropy, two-point functions of the energy-momentum tensor, and holographic $g$-theorem \cite{Miao:2020oey}. Furthermore, it is proved to be equivalent to AdS/CFT for one novel class of solutions \cite{Miao:2020oey}. Unlike the usual doubly holography, wedge holography includes a massless graviton on the EOW branes \cite{Hu:2022lxl}. Remarkably, the effective action on the branes is ghost-free higher derivative gravity plus a CFT \footnote{Since the effective theory on the brane is obtained from Einstein gravity in bulk, it is natural to be ghost-free. The higher derivative gravity generally suffers the ghost problem. The CFTs play an important role in removing the ghost of higher derivative gravity. }, or equivalently, ghost-free dRGT-type multi-gravity \cite{Hu:2022lxl}. See \cite{Miao:2023mui,Ogawa:2022fhy,Aguilar-Gutierrez:2023zoi} for recent developments in wedge holography. 

\begin{figure}[htbp]
\centering
\subfloat[ \ Cone holography \label{sfig:cone}]{
\includegraphics[width=2.6in]{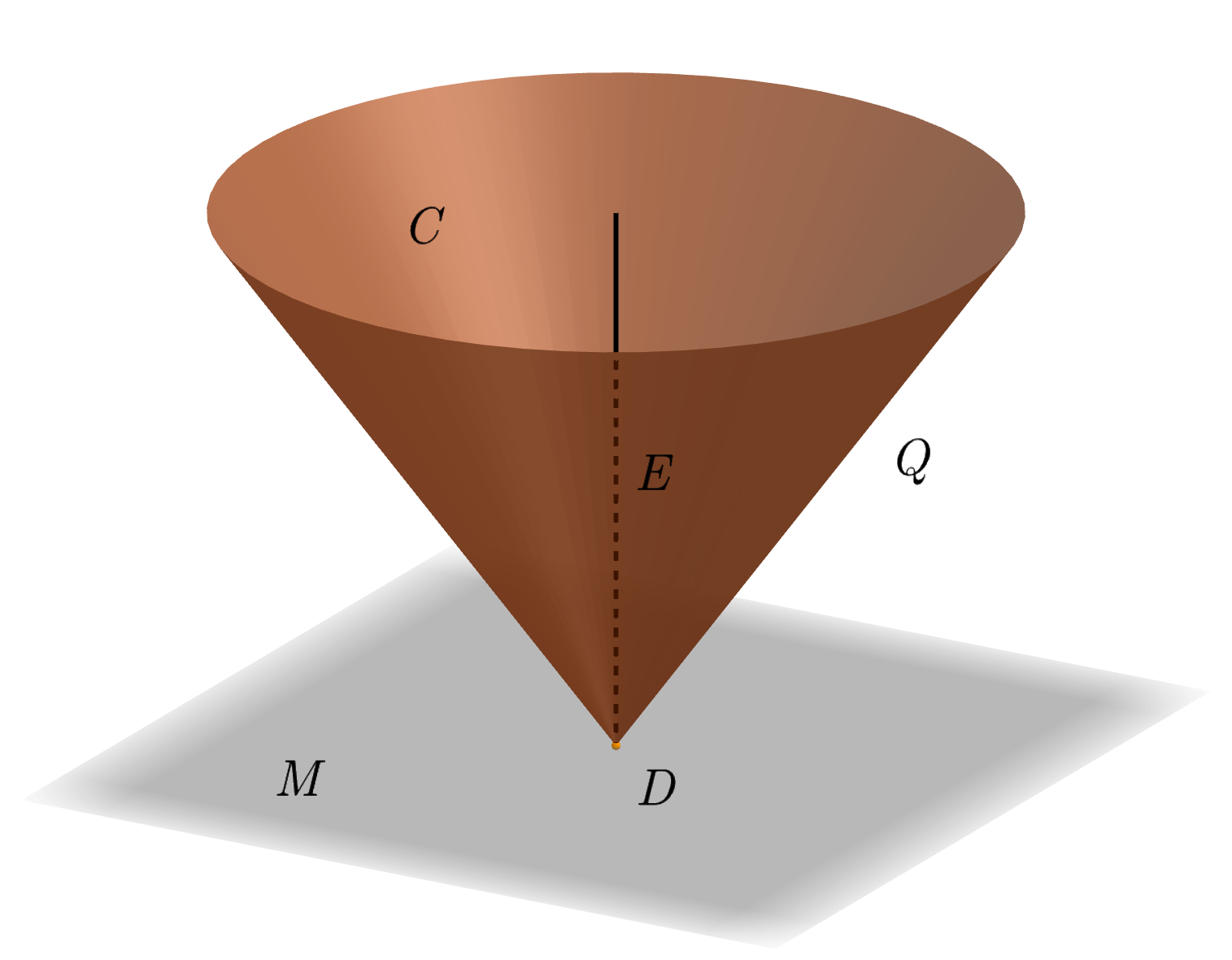}}
\hspace{0.5in}
\subfloat[\ Cone holography from AdS/dCFT  \label{sfig:adscft}]{
\includegraphics[width=2.6in]{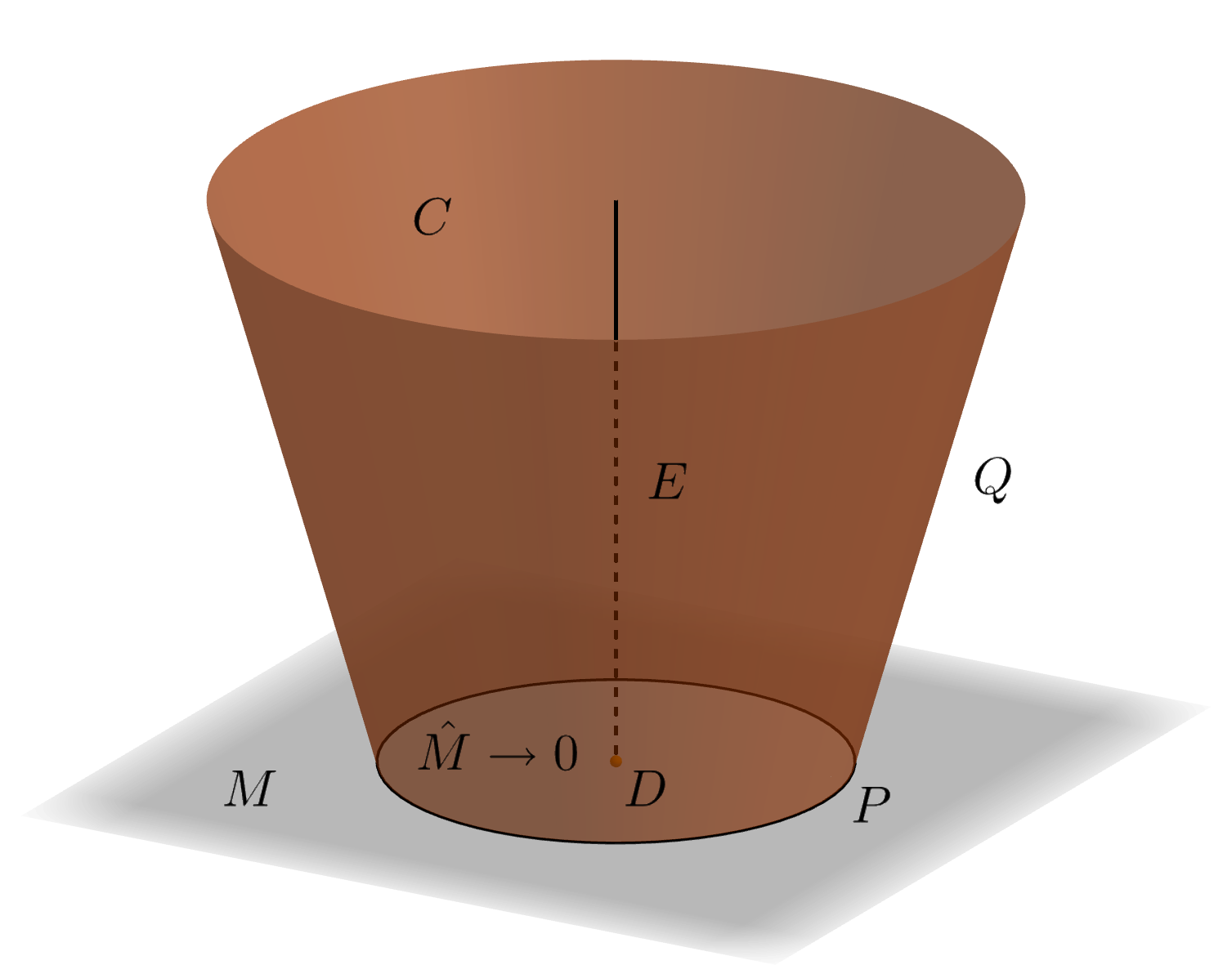}}
\caption{ (left) Geometry of cone holography; (right) Cone holography from AdS/dCFT. In the left figure, $Q$ is a codimension-$1$ brane, $C$ is the cone bounded by $Q$, i.e., $\partial C=Q$, and $E$ (black dotted line) is a codimension-$(q+1)$ brane in bulk. The geometry of $Q$ is set to be $S^q\times \text{AdS}_{p+1}$, and $E$ is set to be AdS$_{p+1}$, so that they shrink to the same defect $D=\partial Q=\partial E$ on the AdS boundary $M$. In the right figure, dCFT lives in the manifold $\hat{M}$ with a boundary $P$ and a codimension-$(q+1)$ defect $D$ at the center. The boundary $P$ and codimension-$(q+1)$ defect $D$ are extended to an EOW brane $Q$ and a codimension-$(q+1)$ brane $E$ in bulk, respectively. In the limit $\hat{M}\to 0$, the bulk spacetime $C$ becomes a cone and we obtain the cone holography from AdS/dCFT.}\label{coneholography}
\end{figure}

Inspired by wedge holography, a codimension-$n$ holography named cone holography \cite{Miao:2021ual} was proposed. See Fig. \ref{coneholography} (left) for the geometry of cone holography. It is conjectured that the classical gravity in $(d+1)$-dimensional conical bulk is dual to the CFT on a $p$-dimensional defect. Similar to wedge holography, cone holography can be derived as a particular limit of AdS/dCFT~\cite{Jensen:2013lxa,DeWolfe:2001pq,Clark:2004sb,Dong:2016fnf}. See Fig. \ref{coneholography} (right) for the schematic. In the zero-volume limit $\hat{M} \to 0$, the bulk modes of dCFT disappear, and only the edge modes on defect $D$ survive. In this sense, cone holography can be regarded as a holographic dual of edge modes on the defect. Cone holography yields the expected Weyl anomaly, entanglement/R\'enyi entropy, two-point functions, and so on \cite{Miao:2021ual}. It also contains a massless graviton on the brane \cite{Miao:2021ual}.  Due to the technical difficulties, the initial model of cone holography \cite{Miao:2021ual} is mainly based on mixed boundary condition (MBC), where one imposes Neumann boundary condition (NBC) on the AdS$_{p+1}$ sector while choosing Dirichlet boundary condition (DBC) on the $S^q$ sector for the EOW brane $Q$. 

The present paper aims to explore cone holography with full NBC on the EOW brane. The NBC is closely related to the junction condition of branes and, thus, is a natural boundary condition. The initial AdS/BCFT is based on NBC~\cite{Takayanagi:2011zk}. Thus, there is good motivation to formulate cone holography with NBC. In general, there are two possible ways to achieve this goal. One is choosing suitable embedding of the EOW brane, which is the way adopted in Ref.~\cite{Miao:2021ual}. However, the embedding function depends on the angle, and it is generally challenging to get analytical solutions. The other method is taking into account the brane-localized field fields. This is the method applied in this paper, used in Ref.~\cite{Kanda:2023zse}. We observe that the obstruction to NBC is due to the asymmetry of the $S^q$ and AdS$_{p+1}$ sectors in the EOW brane $Q=S^q\times \text{AdS}_{p+1}$. Considering the brane-localized $p$-form fields, one can compensate for the asymmetry to satisfy NBC. The brane-localized gravity, such as Dvali-Gabadadze-Porrati (DGP) gravity~\cite{Dvali:2000hr}, can do the same job. However, it requires a negative DGP gravity and thus suffers the ghost problem \cite{Miao:2023mui}. On the other hand, the brane-localized $p$-form field takes the standard form and is ghost-free. Besides, we analyze the perturbative solution and verify that the mass spectrum of Kaluza-Klein (KK) gravitons is non-negative, which strongly supports our model. Furthermore, we prove that our model obeys the holographic $c$-theorem, which is another support to our results. Finally, inspired by the chiral theory in AdS/BCFT \cite{Miao:2022oas,Zhao:2023rno}, we realize NBC for cone holography with a codimension-$2$ tensive brane $E$ by applying brane-localized massive vector fields on the EOW brane $Q$. 

This paper is organized as follows. In section~\ref{sec:review}, we briefly review cone holography and the problem with NBC. We find negative DGP gravity enables NBC but suffers the ghost problem. In section~\ref{sec:toymodel}, we study a toy model of AdS$_4$/CFT$_1$. We find the brane-localized $U(1)$ gauge field can accomplish NBC for cone holography. Section~\ref{sec:coneNBC1} generalizes the toy model to arbitrary dimensions and achieves NBC by brane-localized $p$-form fields. Section~\ref{sec:perturbations} is devoted to analyzing the tensor perturbation and mass spectrum. Section~\ref{sec:ctheorem} discusses the $c$-theorem of cone holography. In section~\ref{sec:coneNBC2}, we construct one class of cone holography with NBC by brane-localized massive vector fields. Finally, we summarize the results and discuss some open problems in section~\ref{sec:conclusions}.

\section{Review of cone holography} \label{sec:review}

This section briefly reviews cone holography and the problem of NBC. Ultimately, we show negative DGP gravity can achieve NBC but has the ghost problem. We leave the resolution to this problem in the following sections. 

\subsection{Cone holography}

Let us recall the main ideas of cone holography. We start with AdS/dCFT shown in Fig. \ref{coneholography} (right). Here, dCFT denotes a CFT coupled with a codimension-$1$ defect $P$ (boundary) and a codimension-$(q+1)$ defect $D$ on the AdS boundary $\hat{M}$. The boundary $P$ and defect $D$ are dual to an EOW brane $Q$ and a codimension-$(q+1)$ brane $E$ in bulk $C$, respectively. AdS/dCFT claims that a dCFT coupled with two defects $D$ and $P$ on the AdS boundary $\hat{M}$ is dual to the gravity coupled with two branes $E$ and $Q$ in bulk $C$. By taking the zero-volume limit $\hat{M} \to 0$, we obtain cone holography as shown in Fig. \ref{coneholography} (left). Since bulk modes in $\hat{M}$ disappear and only the edge modes on defects $P\to D$ survive in such limit, cone holography can be regarded as a holographic dual of edge modes on defects. In cone holography, as shown in Fig. \ref{coneholography} (left), it is proposed that the classical gravity in conical bulk $C$ is dual to ``quantum gravity'' on an EOW brane $Q$ together with a codimension-$(q+1)$ brane $E$, and is dual to CFT on the defect $D=\partial Q=\partial E$. Here, ``quantum gravity'' follows the statement in wedge holography~\cite{Akal:2020wfl} and refers to semi-classical gravity theory containing higher curvature gravity plus quantum matter on a classical background. The notations of cone holography are summarized in Tab. \ref{sect2:conenotation}. Note that the geometry of cone holography is generally non-factorizable. There are coordinate mixing between different sectors $\mathbb{R}^1, S^q, \text{AdS}_{p+1}$. See (\ref{eq:lineeAll}) for an example. However, we still label the geometry by the direct product $\mathbb{R}^1\times S^q\times \text{AdS}_{p+1}$ to emphasize the topology. 

\begin{table}[htb]
\caption{Notations of cone holography}
\centering
    \begin{tabular}{| c | c | c | c |  c | c | c | c| c| c|c| }
    \hline
    & Conical bulk $C$ & EOW brane $Q$  & Brane $E$ & Defect $D$  \\ \hline
  Topology sectors & $\mathbb{R}^1\times S^q\times \text{AdS}_{p+1}$ & $S^q\times \text{AdS}_{p+1}$ & AdS$_{p+1}$ & $\partial E$\\ \hline
 Coordinate & $x^A=(r,x^{\alpha})$ & $x^{\alpha}=(x^a,x^i)$ & $x^{i}=(z,y^{\hat{i}})$ & $y^{\hat{i}}$ \\ \hline
 Dimension & $d+1$ & $d$ & $p+1$ & $p$ \\ \hline
 Metric& $g_{AB}$ & $h_{\alpha\beta}$ & $\gamma_{ij}$ &$\sigma_{\hat{i}\hat{j}}$\\ \hline
   \end{tabular}
\label{sect2:conenotation}
\end{table}

There are three perspectives describing cone holography~\cite{Miao:2021ual}.
\begin{itemize}
 \item \textbf{Perspective 1}: A classical gravity in the conical bulk $C$. Here $C$ is a $(d+1)$-dimensional asymptotically AdS spacetime with the typical metric
        \begin{align}\label{eq:lineeAll}
           \mathrm{d}s_C^2=g_{AB}\mathrm{d}x^A\mathrm{d}x^B=\mathrm{d}r^2+\sinh^2(r)\mathrm{d}\Omega_{q}^2+\cosh^2(r)\mathrm{d}s_{\text{AdS}_{p+1}}^2\,,
        \end{align}
   where the metrics of the unit $q$-sphere and the $(p+1)$-dimensional AdS spacetime respectively are
        \begin{align}
          \mathrm{d}\Omega_{q}^2= & \mathrm{d}\theta^2+\sin^2(\theta)\mathrm{d}\Omega_{q-1}^2\,, \label{eq:sphere}\\
          \mathrm{d}s_{\text{AdS}_{p+1}}^2 = & \frac{\mathrm{d}z^2+\sigma_{\hat{i}\hat{j}}\mathrm{d}y^{\hat{i}}\mathrm{d}y^{\hat{j}}}{z^2}=
          \frac{1}{z^2}\left(\mathrm{d}z^2-\mathrm{d}t^2+\sum_{\hat{i}=1}^{p-1} \mathrm{d}y_{\hat{i}}^2\right)\,. \label{eq:AdS}
        \end{align}
        
 \item \textbf{Perspective 2}: ``Quantum gravity'' on an EOW brane $Q$ and codimension-$(q+1)$ brane $E=\text{AdS}_{p+1}$. The typical EOW brane locates at $r=\rho$, which yields the induced metric
     \begin{align}\label{eq:lineeQ}
        \mathrm{d}s_Q^2=h_{\alpha\beta}\mathrm{d}x^\alpha\mathrm{d}x^\beta=\sinh^2(\rho)
        \mathrm{d}\Omega_{q}^2+\cosh^2(\rho)\mathrm{d}s_{\text{AdS}_{p+1}}^2\,.
     \end{align}
        It should be mentioned that the EOW brane $Q$ is not an AdS brane. The codimension-$(q+1)$ brane $E$ locates at $r=0$ and the induced metric on $E$ is 
     \begin{align}\label{eq:lineeE}
       \mathrm{d}s_E^2=\gamma_{ij}\mathrm{d}x^i\mathrm{d}x^j=\mathrm{d}s_{\text{AdS}_{p+1}}^2\,.
     \end{align}
       The conformal invariance of defect $D$ requires that the brane $E$ is an AdS brane. For convenience, we also employ the following notation
     \begin{align}\label{eq:lineesimple}
      \mathrm{d}s_Q^2=\sinh^2(\rho)\vartheta_{ab} \mathrm{d}x^a\mathrm{d}x^b+\cosh^2(\rho)\gamma_{ij}\mathrm{d}x^i\mathrm{d}x^j\,.
     \end{align}
 \item \textbf{Perspective 3}: A CFT on the $p$-dimensional defect $D=\partial E=\partial Q$. The defect is at $z=0$ and the typical metric is 
     \begin{align}\label{eq:lineeD}
       \mathrm{d}s_D^2=\sigma_{\hat{i}\hat{j}}\mathrm{d}y^{\hat{i}}\mathrm{d}y^{\hat{j}}=-\mathrm{d}t^2+\sum_{\hat{i}=1}^{p-1} \mathrm{d}y_{\hat{i}}^2.
     \end{align}
\end{itemize}

The action of cone holography in bulk takes the form
\begin{align}\label{eq:action}
 I=&\frac{1}{16\pi G_{\text{N}}}\int_C \mathrm{d}^{d+1}x\sqrt{-g}\left(R-2\Lambda\right)+\frac{1}{8\pi G_{\text{N}}}\int_Q\mathrm{d}^dx\sqrt{-h} \left(K-T_Q+L_Q\right) \nonumber \\ &-\frac{1}{8\pi G_{\text{N}}}\int_E\mathrm{d}^{p+1}x\sqrt{-\gamma}\,T_E,
\end{align}
where $G_{\text{N}}$ denotes the gravitational constant, $R$ is the Ricci scalar, $-2\Lambda=d(d-1)$ is the cosmological constant (we have set AdS radius $L=1$), $d=q+p+1$, $K$ labels the extrinsic curvature, $L_Q$ denotes possible brane-localized gravity and matter fields, $T_Q$ and $T_E$ are the tensions of branes $Q$ and $E$, respectively. For simplicity, we set $16\pi G_N=1$ in the following discussions. According to \cite{Bostock:2003cv}, there is no well-defined thin brane limit in Einstein gravity for codimensions higher than two. As a result, we have $T_E=0$ for the codimension-$(q+1)$ brane $E$ with $q\ge 2$. As for the codimension-$2$ brane $E$ with $q=1$, the tension $T_E$ is non-trivial and characterizes the conical singularity
\begin{align}\label{sect2:TE}
T_E=2\pi \left(1-\frac{1}{n}\right),
\end{align}
where $2\pi n$ is the period of angle, i.e., $\theta \sim \theta+2\pi n$.

\subsection{NBC problem}

In the following, we discuss the NBC problem of cone holography. From the action \eqref{eq:action}, we derive NBC on EOW brane $Q$
\begin{align}\label{eq:NBC}
\text{NBC}:\ K_{\alpha\beta}-(K-T_Q) h_{\alpha\beta}=T_{\alpha\beta},
\end{align}
where $T_{\alpha\beta}=\frac{-2}{\sqrt{-h}} \frac{\delta (\sqrt{-h} L_{Q})}{\delta h^{\alpha\beta}}$. We first discuss the minimal model with $T_{\alpha\beta}=L_{Q}=0$ for simplicity. We show below that the typical metric \eqref{eq:lineeAll} does not satisfy NBC on the EOW brane located at $r=\rho$. From NBC \eqref{eq:NBC} with $T_{\alpha\beta}=0$, we derive
\begin{align}\label{sect2:NBC1}
K_{\alpha\beta}=\frac{d}{d-1}T_Q\ h_{\alpha\beta}.
\end{align}
For the typical metric \eqref{eq:lineeAll} in Gauss normal coordinates, the extrinsic curvatures take simple expressions 
\begin{align}\label{eq:EC}
 K_{\alpha\beta}=\frac{1}{2}\frac{\partial h_{\alpha\beta}}{\partial r}\big{|}_{r=\rho},
\end{align}
which yields
\begin{align}\label{sect2:NBC2}
S^q\ \text{sector}:& \  K_{ab}=\coth (\rho ) \ h_{ab}, \\
\text{AdS}_{p+1}\ \text{sector}:& \ K_{ij}=\tanh (\rho ) \ h_{ij}. \label{sect2:NBC3}
\end{align}
Clearly, \eqref{sect2:NBC2} and \eqref{sect2:NBC3} cannot satisfy NBC \eqref{sect2:NBC1} at the same time unless the EOW brane locates at infinity $\rho\to \infty$. This is the NBC problem for cone holography, which comes from the asymmetry of the $S^q$ sector and the AdS$_{p+1}$ sector on the EOW brane. 

There are several possible ways to solve the above problem. First, one chooses mixed boundary condition instead of NBC
\begin{align}\label{sect2:mixed BC}
\text{MBC}:\ \delta h_{ab}=0, \quad K_{ij}-(K-T_Q) h_{ij}=0.
\end{align}
That one imposes DBC on the $S^q$ sector while imposing NBC on the AdS$_{p+1}$ sector on the EOW brane~\cite{Miao:2021ual}. Second, one considers more general embedding functions $r=r(\theta)$. However, it is difficult to find analytical solutions for the case with a tensive brane $E$~\cite{Miao:2021ual}. Besides, the configuration becomes asymmetrical, and the EOW brane $Q$ approaches the AdS boundary $M$ at specific angle $\theta$, which is not the geometry of Fig. \ref{coneholography} (left)~\cite{Miao:2021ual}. Third, one considers more general NBC \eqref{eq:NBC} by adding suitable intrinsic gravity or matter fields on the EOW brane. Let us take DGP gravity with $L_Q=\lambda R_Q$ as an example. The NBC \eqref{eq:NBC} becomes
\begin{align}\label{sect2:DGPNBC}
K_{\alpha\beta}-(K-T_Q) h_{\alpha\beta}=\lambda( R_Qh_{\alpha\beta}-2R_{Q \ \alpha\beta}).
\end{align}

From the induced metric \eqref{eq:lineeQ}, we derive intrinsic curvatures on the EOW brane located at $r=\rho$,
\begin{align}\label{sect2:DGPR}
R_{Q\ ab}=(q-1) \text{csch}^2(\rho ) h_{ab}, \quad R_{Q\ ij}=-p\ \text{sech}^2(\rho ) h_{ij}.
\end{align}
Interestingly, the above curvature asymmetry can compensate those of extrinsic curvatures \eqref{sect2:NBC2} and \eqref{sect2:NBC3}, which enables the NBC \eqref{sect2:DGPNBC} for suitable parameters $T_Q$ and $\lambda$. Substituting the intrinsic curvatures \eqref{sect2:DGPR} together with extrinsic curvatures \eqref{sect2:NBC2} and \eqref{sect2:NBC3} into NBC \eqref{sect2:DGPNBC}, we derive a negative DGP parameter
\begin{align}\label{sect2:lambda}
\lambda =\frac{1}{-2 p \tanh (\rho )-2 (q-1) \coth (\rho )} <0.
\end{align}
 We do not show the tension $T_Q$ since it is irrelevant to our discussions. According to \cite{Miao:2023mui}, negative DGP gravity has the ghost problem. Thus, although negative DGP gravity can solve the NBC problem, it is not well-defined. The following sections will resolve the NBC problem by applying brane-localized gauge fields.

In the above discussions, we mainly focus on the AdS space \eqref{eq:lineeAll} in bulk. It is no longer a solution for a tensive brane $E$ due to the back-reaction. As discussed above, no thin brane can consistently couple with Einstein gravity for codimensions higher than two. Thus, the only non-trivial brane $E$ is codimension-$2$. For a tensive codimension-$2$ brane $E$, the bulk metric is given by 
\begin{align}\label{rbarmetric}
\mathrm{d}s^2=\frac{\mathrm{d}\bar{r}^2}{f(\bar{r})}+ f(\bar{r}) \mathrm{d}\theta^2+\bar{r}^2 \frac{\mathrm{d}z^2-\mathrm{d}t^2+\sum_{\hat{i}=1}^{p-1}\mathrm{d}y_{\hat{i}}^2}{z^2},
\end{align}
with
\begin{align}\label{rbarF}
f(\bar{r})=\bar{r}^2-1-\frac{\bar{r}_h^{d-2}(\bar{r}_h^{2}-1)}{\bar{r}^{d-2}}\,, \quad \bar{r}_h=\frac{1+\sqrt{d^2 n^2-2 d n^2+1}}{d n}\,,
\end{align}
where the codimension-$2$ brane $E$ locates at $\bar{r}=\bar{r}_h$. Note that $2\pi n$ is the period of angle $\theta$, and is related to the brane tension \eqref{sect2:TE}. The tension-less case corresponds to $n=\bar{r}_h=1$. Performing the radial coordinate transformation
\begin{align}\label{rbarr}
\mathrm{d}r=\frac{\mathrm{d}\bar{r}}{\sqrt{f(\bar{r})}},
\end{align}
one can rewrite the metric \eqref{rbarmetric} into the form of \eqref{eq:lineeAll}.

In summary, this section reviews cone holography and the NBC problem. We show that negative DGP gravity can resolve the NBC problem but suffers the ghost problem.

\section{A toy model with brane-localized Maxwell fields}
\label{sec:toymodel}

In this section, we study a toy model of AdS$_4$/CFT$_1$, the simplest cone holography. We show that Maxwell fields on the EOW brane can make consistent NBC.

In AdS$_4$/CFT$_1$, the line element~\eqref{eq:lineeAll} becomes
\begin{align}\label{eq:lineetoy}
  \mathrm{d}s^2=\mathrm{d}r^2+\sinh^2(r)\mathrm{d}\theta^2+\cosh^2(r)\frac{\mathrm{d}z^2-\mathrm{d}t^2}{z^2},
\end{align}
where the codimension-$2$ brane $E$ and EOW brane $Q$ are at $r=0$ and $r=\rho$, respectively.  

Considering an intrinsic $U(1)$ gauge field $A_{\alpha}$ on the EOW brane $Q$. The Lagrangian is given by
\begin{align}\label{eq:Lagrangian-Maxwell}
  L_Q=-\frac{1}{4}F^{\alpha\beta}F_{\alpha\beta},
\end{align}
where $F=\mathrm{d}A$ denotes the field strength. The Lagrangian \eqref{eq:Lagrangian-Maxwell} yields the Maxwell equation
\begin{align}\label{eq:EOM-Maxwell}
  \nabla_\alpha F^{\alpha\beta}=& 0\,,
\end{align}
and the energy-momentum tensor on the EOW brane
\begin{align}\label{eq:EMT-Maxwell}
  T_{\alpha\beta}=F_{\alpha \sigma} F_\beta^{\;\sigma}- \frac{1}{4} h_{\alpha\beta} F_{\lambda\sigma} F^{\lambda\sigma}.
\end{align}
Then NBC \eqref{eq:NBC} becomes
\begin{align}\label{sect3:NBC}
K_{\alpha\beta}-(K-T_Q) h_{\alpha\beta}=F_{\alpha \sigma} F_\beta^{\;\sigma}- \frac{1}{4} h_{\alpha\beta} F_{\lambda\sigma} F^{\lambda\sigma}.
\end{align}

We make the following ansatz of brane-localized Maxwell fields
\begin{align}\label{eq:vectorA}
 A_{\alpha}=\left(0,0,A_{t}(z) \right).
\end{align}
Solving the Maxwell equation \eqref{eq:EOM-Maxwell}, we get
\begin{align}\label{sect3: vector}
A_{t}(z)=c_0+\frac{c_1}{z},
\end{align}
where $c_0, c_1$ are integration constants. We remark that $c_1$ can be interpreted as the charge in the usual AdS$_2$/CFT$_1$. Substituting \eqref{eq:vectorA} and \eqref{sect3: vector} into \eqref{eq:EMT-Maxwell}, we obtain
\begin{align}\label{sect3:Tij}
 T_{\theta\theta}=\frac{1}{2} c_1^2\ \text{sech}^4(\rho ) h_{\theta\theta}, \quad T_{ij}=-\frac{1}{2} c_1^2\ \text{sech}^4(\rho ) h_{ij},
\end{align}
where $\left(i,j\right)$ denote the $\left(z,t\right)$ components and $h_{\theta\theta}=\sinh^2(\rho)$, $h_{zz}=-h_{tt}=\cosh^2(\rho)/z^2$. Similar to the DGP gravity of section~\ref{sec:review}, the asymmetry of energy-momentum tensor \eqref{sect3:Tij} can compensate for the asymmetry of extrinsic curvature \eqref{sect2:NBC2} and \eqref{sect2:NBC3} to satisfy the NBC \eqref{sect3:NBC}. One can easily check that the NBC \eqref{sect3:NBC} can be satisfied provided the following parameters are chosen 
\begin{align}\label{sect3: parameter}
c_1^2=\cosh ^2(\rho ) \coth (\rho ), \quad T_Q=\tanh (\rho )+\coth (\rho )-\text{csch}(2 \rho ).
\end{align}

Some comments are in order. 
\begin{itemize}
  \item First, similar to DGP gravity, $U(1)$ gauge field can also accomplish NBC on the EOW brane $Q$.
  \item Second, better than DGP gravity, the brane-localized $U(1)$ gauge field \eqref{eq:Lagrangian-Maxwell} has the correct sign and thus is ghost-free.
  \item Third, one do not need to worry about the divergence of gauge field on the defect $z=0$ in \eqref{sect3: vector}. Recall that the asymptotic solution of Maxwell fields takes the following form near the AdS$_{p+1}$ boundary
\begin{align}\label{sect3:asymptotic solution}
A(y) \sim c_0(y) z^0+ c_1(y) z^{p-2}+\cdots, \ \text{for} \ z\sim 0,
\end{align}
where $c_0(y) $ is the background field and  $c_1(y)$ denotes the charge/current density on the AdS boundary. 
Thus, \eqref{sect3: vector} takes the standard expression in AdS$_2$/CFT$_1$ and nothing goes wrong (recall $E$ is an AdS$_2$ brane with $p=1$). Note that $c_1$ defined in the vector \eqref{sect3: vector} is the electric charge. It means we need charged branes to satisfy NBC.
  
 \item Fourth, the key point to achieve NBC is to compensate the asymmetry of the $S^q$ and AdS$_{p+1}$ sectors on the EOW brane $Q$. We find that $p$-form gauge fields can do the job for cone holography in general dimensions.
\end{itemize}

\section{Cone holography with NBC I}
\label{sec:coneNBC1}

This section extends the results of section~\ref{sec:toymodel} to the cone holography with general dimensions and codimensions. As discussed above, the critical point is to compensate for the asymmetry of the two sectors of EOW brane $Q$ by employing suitable brane-localized fields. Besides, we require the theory to be well-defined and ghost-free. We find $p$-form fields on the EOW brane can achieve this goal. In this section, we treat the cases of tensionless and tensive $E$ branes in a unified way.

The line element~\eqref{eq:lineeAll} of cone holography can be generalized to be~\cite{Miao:2021ual}
\begin{align}\label{eq:lineeg}
 \mathrm{d}s^2=\mathrm{d}r^2+b(r)^2\vartheta_{ab}(x^c)\mathrm{d}x^ax^b+a(r)^2\gamma_{ij}(x^k)\mathrm{d}x^i \mathrm{d}x^j,
\end{align}
where $a(r)$ and $b(r)$ are warp factors determined by Einstein equations, $\vartheta_{ab}$ and $\gamma_{ij}$ are metrics of the unite $q$-sphere and the AdS$_{p+1}$, respectively. For tensionless brane $E$, we have $b(r)=\sinh(r)$ and $a(r)=\cosh(r)$. While for tensive brane $E$, there are no analytical expressions of $b(r)$ and $a(r)$ generally \footnote{In principle, one can perform the coordinate transformation \eqref{rbarr} to derive $b(r), a(r)$ from metric \eqref{rbarmetric} for tensive codimension-$2$ brane $E$. However, in general, the integral in the coordinate transformation \eqref{rbarr} cannot be worked out exactly.}. Fortunately, we do not need exact $b(r)$ and $a(r)$ for our purpose. From \eqref{eq:lineeg}, we derive the extrinsic curvatures on the EOW brane ($r=\rho$)
\begin{align}\label{eq:EC-componentsg}
 K_{ab} =& \frac{1}{2}\frac{\partial h_{ab}}{\partial r}|_{r=\rho}=b(\rho) b'(\rho)\vartheta_{ab}=\frac{b'(\rho)}{b(\rho)}h_{ab}\,, \\
 K_{ij} =& \frac{1}{2}\frac{\partial h_{ij}}{\partial r}|_{r=\rho}=a(\rho) a'(\rho)\gamma_{ij} =\frac{a'(\rho)}{a(\rho)}h_{ij}\,. \label{eq:EC-componentsgij}
\end{align}
where the ${}'$ denotes the derivative with respect to $\rho$. The trace of extrinsic curvature $K_{\alpha\beta}$ is given by
\begin{align}\label{eq:ECg}
 K =& h^{ab}K_{ab}+h^{ij}K_{ij} = q \frac{b'(\rho)}{b(\rho)}+(p+1)\frac{a'(\rho)}{a(\rho)} \,.
\end{align}
As mentioned in section~\ref{sec:review}, the asymmetry of extrinsic curvatures \eqref{eq:EC-componentsg} and \eqref{eq:EC-componentsgij} is the central obstruction to NBC.

Consider $p$-form gauge fields on the EOW brane with the Lagrangian density
\begin{align}\label{eq:Lagrangian-pform-compact}
 L_Q= -\frac{1}{2(p+1)!}F^{\mu_1\mu_2\cdots\mu_{p+1}}F_{\mu_1\mu_2\cdots\mu_{p+1}},
\end{align}
where the $p$-form field strength $F_{\mu_1\mu_2\cdots\mu_{p+1}}$ reads
\begin{align}\label{eq:field-strength}
 F_{\mu_1\mu_2\cdots\mu_{p+1}}=(p+1)\partial_{[\mu_1}A_{\mu_2\cdots\mu_{p+1}]}\,.
\end{align}
Similar to the $U(1)$ gauge fields, the $p$-form field strength \eqref{eq:field-strength} is invariant under the gauge transformation
\begin{align}\label{eq:gauge-trans}
 A_{\mu_1\mu_2\cdots\mu_{p}}\rightarrow A_{\mu_1\mu_2\cdots\mu_{p}}+p\partial_{[\mu_1} \Theta_{\mu_2\cdots\mu_{p]}},
\end{align}
where $\Theta_{\mu_2\cdots\mu_{p}}$ is a $(p-1)$-form gauge function. Similarly, the $p$-form field satisfies the Bianchi identity
\begin{align}\label{eq:Bianchi-pform}
 \nabla_{[\mu_1}F_{\mu_2\cdots\mu_{p+2}]}=0\, ,
\end{align}
and the equation of motion
\begin{align}\label{eq:EOM-pform}
 \nabla_{\mu_1}F^{\mu_1\mu_2\cdots\mu_{p+1}}=0\,,
\end{align}
which governs the electrodynamics of the $p$-form field.

From the Lagrangian \eqref{eq:Lagrangian-pform-compact}, we obtain the symmetric and conserved energy-momentum tensor of $p$-form as
\begin{align}\label{eq:EMT-pform}
 T_{\alpha\beta}=&-2\frac{\partial L_Q}{\partial h^{\alpha\beta}}+h_{\alpha\beta}L_Q \nonumber \\
 =&\frac{1}{p!}F_{\alpha\,\mu_1\mu_2\cdots\mu_{p}} F_\beta^{\ \mu_1\mu_2\cdots\mu_{p}}- \frac{1}{2(p+1)!} h_{\alpha\beta} F_{\mu_1\mu_2\cdots\mu_{p+1}} F^{\mu_1\mu_2\cdots\mu_{p+1}}\,.
\end{align}
For the sake of simplicity, we choose the $p$-form field with only one non-zero component $A_{ty_1y_2\cdots y_{p-1}}(z)$. After some calculations, we derive the energy-momentum tensor
\begin{align}\label{eq:EMTgab}
 T_{ab}=&\frac{1}{2}b(\rho)^2\left(\frac{z^2}{a(\rho)^2}\right)^{p+1} \left[\partial_z A_{ty_1y_2\cdots y_{p-1}}(z)\right]^2 \vartheta_{ab}\,\nonumber \\
 =&\frac{1}{2}\left(\frac{z^2}{a(\rho)^2}\right)^{p+1} \left[\partial_z A_{ty_1y_2\cdots y_{p-1}}(z)\right]^2 h_{ab},\\
 T_{ij}=&-\frac{1}{2} a(\rho)^2 \left(\frac{z^2}{a(\rho)^2}\right)^{p+1} \left[\partial_z A_{ty_1y_2\cdots y_{p-1}}(z)\right]^2 \gamma_{ij} \nonumber \\
 =& -\frac{1}{2} \left(\frac{z^2}{a(\rho)^2}\right)^{p+1} \left[\partial_z A_{ty_1y_2\cdots y_{p-1}}(z)\right]^2 h_{ij}. \label{eq:EMTgij}
\end{align}
which takes the exact forms to compensate the asymmetry of extrinsic curvatures \eqref{eq:EC-componentsg} and \eqref{eq:EC-componentsgij} in the NBC \eqref{eq:NBC}, provided $A_{ty_1y_2\cdots y_{p-1}}(z)\sim 1/z^p$. Remarkably, the equation of motion \eqref{eq:EOM-pform} indeed gives $A_{ty_1y_2\cdots y_{p-1}}(z)\sim 1/z^p$. That is one of the reasons we choose $p$-form fields instead of other form fields. Only the $p$-form field can make everything consistent. 

Solving the equation of motion \eqref{eq:EOM-pform} and NBC \eqref{eq:NBC}, we finally obtain 
\begin{align}\label{eq:solutiong}
  A_{ty_1y_2\cdots y_{p-1}}(z)=&\pm \frac{a(\rho)^p}{p\, z^p}a(\rho)\sqrt{\frac{b'(\rho)}{b(\rho)}-\frac{a'(\rho)}{a(\rho)}}\,, \nonumber\\
  T_Q=&\left(q-\frac{1}{2}\right) \frac{b'(\rho)}{b(\rho)}+\left(p+\frac{1}{2}\right)\frac{a'(\rho)}{a(\rho)}\,.
\end{align}
It is worth mentioning that the condition 
\begin{align}\label{eq:constraintg}
 \frac{b'(\rho)}{b(\rho)}-\frac{a'(\rho)}{a(\rho)}>0\,,
\end{align}
should be satisfied for a real $p$-form field. For a tensionless brane $E$, $a(r)=\cosh(r)$ and $b(r)=\sinh(r)$ indeed meet the constraint~\eqref{eq:constraintg}. After replacing the warp factors $a(\rho)=\cosh(\rho)$ and $b(\rho)=\sinh(\rho)$, the solution becomes
\begin{align}\label{eq:solutionS}
 A_{ty_1y_2\cdots y_{p-1}}(z)=&\pm \frac{\cosh ^p(\rho )}{p\, z^p}\sqrt{\coth (\rho )}\,, \nonumber\\
 T_Q=& q \coth (\rho )+p\tanh (\rho )-\text{csch}(2 \rho )\,,
\end{align}
which agrees with the toy model with $p=q=1$.  Recall only the codimension-$2$ brane $E$ can have non-trivial tension.  Comparing \eqref{rbarmetric} with \eqref{eq:lineeg} for this case, we get $b(r)=\sqrt{f(\bar{r})}$, $a(r)=\bar{r}$ and $\mathrm{d}r=\mathrm{d}\bar{r}/\sqrt{f(\bar{r})}$. Then we have
\begin{align}\label{eq:constraintgtensive}
 \frac{b'(\rho)}{b(\rho)}-\frac{a'(\rho)}{a(\rho)}=&\frac{\mathrm{d} \sqrt{f(\bar{r})}}{\mathrm{d}\bar{r}}-\frac{ \sqrt{f(\bar{r})} }{\bar{r}}\nonumber\\
=&\frac{d \ \bar{r}^{2-d} \left(\bar{r}_h^2-1\right) \bar{r}_h^{d-2}+2}{2 \bar{r} \sqrt{f\left(\bar{r}\right)}}>\frac{d \left(\bar{r}_h^2-1\right)+2}{2 \bar{r} \sqrt{f\left(\bar{r}\right)}}=\frac{ \bar{r}_h}{n \bar{r} \sqrt{f\left(\bar{r}\right)}}>0,
\end{align}
where we have used
\begin{align}\label{eq:conditions}
  \bar{r}> \bar{r}_h\,, \quad f(\bar{r})>0\,, \quad \bar{r}_h=\frac{1+\sqrt{d^2 n^2-2 d n^2+1}}{d n}\le 1
\end{align}
above. Note that the above $\bar{r}$ takes value on the EOW brane $Q$, i.e., $\bar{r}=\bar{\rho}> \bar{r}_h$. Now we have verified the condition \eqref{eq:constraintg} for tensive brane $E$ too.

In summary, we have successfully constructed cone holography with NBC by using $p$-form fields on the EOW brane. The $p$-form field is ghost-free and takes the standard form \eqref{eq:Lagrangian-pform-compact}. Thus, this model is well-defined.

\section{Tensor perturbation analyses}
\label{sec:perturbations}

This section investigates the linear tensor perturbations of the metric and $p$-form fields. For our purpose, we focus on the transverse-traceless (TT) tensor perturbations on the codim-$(q+1)$ brane $E$, or equivalently, the AdS$_{p+1}$ sector of EOW brane $Q$. Interestingly, the tensor mode decouples from the perturbations of $p$-form fields in NBC. As a result, the mass spectrum of KK gravitons on $E$ reduces to that of mixed boundary condition (MBC)~\cite{Miao:2021ual}\footnote{Recall that MBC imposes NBC on the AdS$_{p+1}$ sector and DBC on the $S^q$ sector. Since we focus on the AdS$_{p+1}$ sector and the $p$-form field decouples, the mass spectrum of KK gravitons on the AdS$_{p+1}$ sector are the same for MBC and NBC.}. The initial work \cite{Miao:2021ual} has analyzed the mass spectrum of KK gravitons for a tensionless brane $E$ in the $s$-wave case (no angle dependences in metric perturbations). Here, we discuss the most general cases and prove that the mass spectrum of KK gravitons is non-negative, which strongly supports the cone holography with both NBC and MBC. 

\subsection{Linear tensor perturbations}

The metric and the $p$-form field can separated into their background and perturbation parts:
\begin{align}\label{eq:general-perturbations}
  \tilde{g}_{AB}= & g_{AB}+\delta g_{AB}\,,\\
  \tilde{A}_{\mu_1\mu_2\cdots\mu_{p}} =& A_{\mu_1\mu_2\cdots\mu_{p}}+\delta A_{\mu_1\mu_2\cdots\mu_{p}}\,.
\end{align}
We are interested in the tensor perturbation as
\begin{align}\label{eq:perturbation}
  \mathrm{d}s^2=\mathrm{d}r^2+b(r)^2\vartheta_{ab}\mathrm{d}x^ax^b+a(r)^2(\gamma_{ij}+H_{ij})\mathrm{d}x^i \mathrm{d}x^j\,,
\end{align}
where the perturbation $H_{ij}$ satisfies the TT gauge
\begin{align}\label{eq:TTcondition}
  \nabla^{(\gamma)}_i H^{ij}=0\,, \quad \gamma^{ij}H_{ij}=0\,,
\end{align}
with $\nabla^{(\gamma)}$ the covariant derivative defined by $\gamma_{ij}$. Recall that $\vartheta_{ab}$ and $\gamma_{ij}$ denote the background metrics of $S^q$ and AdS$_{p+1}$ with unite radiuses, respectively.  Thus, we have
\begin{align}\label{eq:Ricci}
  R^{(\vartheta)}_{ab}=(q-1) \vartheta_{ab}\,,  \quad  R^{(\gamma)}_{ij}=-p \gamma_{ij}\,.
\end{align}
Here, $R^{(\vartheta)}_{ab}$ and $R^{(\gamma)}_{ij}$ are constructed by the metrics $\vartheta_{ab}$ and $\gamma_{ij}$, respectively. From the above equations and Einstein equations in bulk, we obtain the equation of motion of the perturbation $H_{ij}$ (see Appendix~\ref{app:b}) as
\begin{align}\label{eq:EOM_p}
  \frac{1}{a^2}\left(\square^{(\gamma)}+2\right)H_{ij}+ \frac{1}{b^2}\Delta^{(\vartheta)}H_{ij}+\partial_r\partial_r H_{ij}+
  \left[(p+1)\frac{a'}{a}+q\frac{b'}{b}\right]\partial_r H_{ij} =0\,,
\end{align}
where $\square^{(\gamma)}=\gamma^{ij}\nabla^{(\gamma)}_i\nabla^{(\gamma)}_j$ and $\Delta^{(\vartheta)}=\vartheta^{ab}\nabla^{(\vartheta)}_{a}\nabla^{(\vartheta)}_{b}$ are respectively d'Alembert operators in the AdS$_{p+1}$ and the unit sphere $S^q$, and the ${}'$ denotes the derivative with respective to the coordinate $r$.

Next, we perform the standard separation of variables
\begin{align}\label{KKdecomr}
 H_{ij}(r,x^a,x^i)=\psi\left(r\right)\xi(x^a)\chi_{ij}\left(x^{i}\right)\,,
\end{align}
where $\xi(x^a)$ are spherical harmonic functions obeying 
\begin{equation}\label{eq:KG2}
\left[\Delta^{(\vartheta)}+l(l+q-1)\right]\xi(x^a)=0\,,
\end{equation}
and $\chi_{ij}\left(x^{i}\right)$ satisfy the equation of motion of massive gravitons on AdS$_{p+1}$
\begin{equation}\label{eq:KG1}
\left(\square^{(\gamma)}+2-m^2\right)\chi_{ij}\left(x^{i}\right)=0\,.
\end{equation}
Here $m$ denotes the graviton mass, and the $l\ge 0$ are non-negative integers for $q>1$. For $q=1$, $l$ can be non-negative real numbers, and there is a conical singularity for non-integer $l$.
Substituting Eqs.~\eqref{KKdecomr}, \eqref{eq:KG2}, and \eqref{eq:KG1} into the main perturbation equation (\ref{eq:EOM_p}), we get the equation of motion of $\psi\left(r\right)$
\begin{align}\label{eq:EOM_psi}
 \left[ \partial_r\partial_r +
 \left( \left(p+1\right)\frac{a'}{a}+q\frac{b'}{b}\right)\partial_r+\frac{m^2}{a^2}- \frac{l(l+q-1)}{b^2} \right]\psi(r) =0\,.
\end{align}

Above, we have discussed the constraints on metric perturbations from Einstein equations in bulk. Let us continue to study the effects of NBC~\eqref{eq:NBC} on the EOW brane. At the linear orders, the $(a,b)$ and $(i,j)$ components of NBC~\eqref{eq:NBC} read
\begin{subequations}\label{eq:EBDY}
\begin{align}
 \delta K_{ab}-(K-T_Q)\delta h_{ab}-\delta K h_{ab}=&\delta T_{ab}\,,\label{seq:ab}\\
 \delta K_{ij}-(K-T_Q)\delta h_{ij}-\delta K h_{ij}=&\delta T_{ij}\,\label{seq:ij},
\end{align}
\end{subequations}
where $T_{\alpha\beta}$ are the stress tensors (\ref{eq:EMT-pform}) of $p$-form fields. From the metric ansatz (\ref{eq:perturbation}), the TT gauge (\ref{eq:TTcondition}) and the formula of extrinsic curvatures $K_{\alpha\beta}=\partial_r h_{\alpha\beta}/2$, we derive
\begin{eqnarray}\label{sect5:delta ab}
\delta K_{ab}=\delta K=\delta h_{ab}=0,
\end{eqnarray}
which together with (\ref{seq:ab}) yields $\delta T_{ab}=0$. Recall that only $F_{i_1i_2\cdots i_{p+1}}$ is non-zero for the background solution, which gives
\begin{eqnarray}\label{sect5:delta ab1}
\delta F^2&=&2 F^{i_1i_2\cdots i_{p+1}} \delta F_{i_1i_2\cdots i_{p+1}} + (p+1) \delta h^{i_1j_1} F_{i_1i_2\cdots i_{p+1}} F_{j_1}^{\ i_2\cdots i_{p+1}} \nonumber\\
&=&2 F^{i_1i_2\cdots i_{p+1}} \delta F_{i_1i_2\cdots i_{p+1}} + \delta h^{i_1j_1} h_{i_1j_1} F^2\nonumber\\
&=&2 F^{i_1i_2\cdots i_{p+1}} \delta F_{i_1i_2\cdots i_{p+1}}. 
\end{eqnarray}
From (\ref{eq:EMT-pform}) and $\delta T_{ab}=0$, we derive
\begin{eqnarray}\label{sect5:delta ab2}
\delta T_{ab}&=&-\frac{1}{2(p+1)!} h_{ab} \delta F^2=0\nonumber\\
&=& -\frac{1}{(p+1)!} h_{ab} F^{i_1i_2\cdots i_{p+1}} \delta F_{i_1i_2\cdots i_{p+1}}=0,
\end{eqnarray}
which yields for $p$-form fields
\begin{eqnarray}\label{sect5:delta ab3}
\delta F_{i_1i_2\cdots i_{p+1}}=0.
\end{eqnarray}
Above we have used the fact that, for $p$-form fields in $(p+1)$ dimensions, the field strength $F_{i_1i_2\cdots i_{p+1}}$ has only one independent component. 

Let us continue to consider the $(i,j)$ components of NBC~\eqref{eq:NBC}. After some calculations, we obtain
\begin{align}\label{eq:deltaKij}
 \delta K_{ij}=&\frac{1}{2}a^2H'_{ij}+aa'H_{ij}=\frac{1}{2}a^2H'_{ij}+\frac{a'}{a}\delta h_{ij}\,,\\
 \delta T_{ij}=&\frac{1}{2(p+1)!}F^2a^2H_{ij}=\frac{1}{2(p+1)!}F^2\delta h_{ij}\,, \label{eq:deltaTij}
\end{align}
Taking the trace of background NBC~\eqref{eq:NBC} for $(i,j)$ components, we derive
\begin{eqnarray}\label{sect5:traceTij}
h^{ij}\left( K_{ij} -\left(K-T_Q\right) h_{ij} -T_{ij}\right)= (p+1) \frac{a'}{a}-(p+1) \left(K-T_Q\right)-\frac{1}{2 p!} F^2=0. 
\end{eqnarray}
From Eqs. \eqref{seq:ij}, \eqref{eq:deltaKij}, \eqref{eq:deltaTij}, and \eqref{sect5:traceTij}, we obtain 
\begin{align}\label{sect5:NBCHonQ}
 H'_{ij}|_Q=0\,.
\end{align}
Following \cite{Miao:2021ual}, we impose the natural boundary condition on the brane $E$
\begin{align}\label{sect5:NBCHonE}
 H_{ij}|_E \ \text{is\ finite}.
\end{align}
Equivalently, from Eq. (\ref{KKdecomr}), we have
\begin{align}\label{sect5:NBCpsionQ}
 &\psi'|_Q= \psi'(\rho)=0, \\
 &\psi|_E= \psi(0) \ \text{is\ finite}. \label{sect5:NBCpsionE}
\end{align}

Interestingly, the $p$-form field does not affect the boundary condition of the metric perturbation $H_{ij}$ on the AdS$_{p+1}$ sector. As a result, the equation of motion \eqref{eq:EOM_p} and boundary conditions \eqref{sect5:NBCHonQ} and \eqref{sect5:NBCHonE} of $H_{ij}$ are the same for NBC and MBC \cite{Miao:2021ual}, which produces the same mass spectrum of KK gravitons on the AdS$_{p+1}$ sector. Note that MBC fixes the metric while NBC allows the perturbations in the $S^q$ sector. Thus, they are different boundary conditions generally. We leave the study of metric perturbations in the $S^q$ sector to future works. 

\subsection{Mass spectrum of KK gravitons}

This subsection investigates the mass spectrum of KK gravitons on the brane $E$ (AdS$_{p+1}$ sector of EOW brane). As mentioned above, the results apply to both NBC and MBC. Recall that \cite{Miao:2021ual} has studied the mass spectrum on a tensionless brane $E$ for the $s$-wave with $l=0$. The mass spectrum of KK gravitons is non-negative and includes a massless mode. Here, we discuss the most general case. For our purpose, we aim to prove that the mass spectrum of KK gravitons is real and non-negative generally, i.e., $m^2\ge 0$, so that cone holography with NBC/MBC is well-defined. 

From the equation of motion (\ref{eq:EOM_p}) and boundary conditions \eqref{sect5:NBCpsionQ} and \eqref{sect5:NBCpsionE}, we can determine the mass spectrum. For more details, see \cite{Miao:2021ual} and \cite{Hu:2022zgy} for the tensionless and the tensive brane $E$, respectively. In this paper, we are not interested in the exact value of the mass. Instead, we aim to prove $m^2\ge 0$, so we do not need to solve the equation of motion (\ref{eq:EOM_p}) and boundary conditions \eqref{sect5:NBCpsionQ} and \eqref{sect5:NBCpsionE}. 

First, let us prove $m^2$ is real. Note that $m^2$ is just a parameter separating variables (\ref{KKdecomr}). In principle, it can be complex. For the boundary conditions \eqref{sect5:NBCpsionQ} and \eqref{sect5:NBCpsionE}, the orthogonal relationship for KK gravitons reads \footnote{The orthogonal relationship for AdS/BCFT and wedge holography with $q=1, p=d-1$ is given by \cite{Miao:2023mui}, which can be easily generalized to the current case of cone holography. }

\begin{eqnarray}\label{sect5:orthogonal relationship}
\int_0^{\rho} b(r)^q a(r)^{p-1} \psi_{m,l}(r) \psi_{m',l}(r) \mathrm{d}r= c_{l} \delta_{m,m'},
\end{eqnarray}
where $l$ is the real non-negative angle quantum number, $c_l$ is a constant depending on the normalization of $\psi_{m,l}$. Suppose that there are complex $m^2$ in the mass spectrum of KK gravitons. Since the equation of motion (\ref{eq:EOM_p}) and boundary conditions \eqref{sect5:NBCpsionQ} and \eqref{sect5:NBCpsionE} are both real, complex $m^2$ must appear in a complex conjugate pair. By applying the orthogonal condition (\ref{sect5:orthogonal relationship}) for the complex conjugate pair, we get 
\begin{eqnarray}\label{sect5:orthogonal relationship1}
\int_0^{\rho} b(r)^q a(r)^{p-1} \psi_{m,l}(r) \psi_{m^*,l}(r) \mathrm{d}r= 0.
\end{eqnarray}
On the other hand, we have 
\begin{eqnarray}\label{sect5:orthogonal relationship2}
\int_0^{\rho} b(r)^q a(r)^{p-1} \psi_{m,l}(r) \psi_{m^*,l}(r) \mathrm{d}r= \int_0^{\rho} b(r)^q a(r)^{p-1} |\psi_{m,l}(r)|^2 \mathrm{d}r>0,
\end{eqnarray}
where $a(r)$ and $b(r)$ are real positive functions. The contradiction between the expressions \eqref{sect5:orthogonal relationship1} and \eqref{sect5:orthogonal relationship2} suggests that there is no complex mass. Now, we have finished the proof that the mass spectrum of KK gravitons is real. 

Second, let us prove that the mass spectrum of KK gravitons is non-negative, i.e., $m^2\ge 0$. To do so, we construct a non-negative integral 
\begin{eqnarray}\label{sect5:orthogonal relationship3}
\int_0^{\rho}  b(r)^q a(r)^{p+1} \psi_{m,l}'(r) \psi'_{m,l}(r) \mathrm{d}r\ge 0.
\end{eqnarray}
Integrating by parts and applying the equation of motion \eqref{eq:EOM_p} and boundary conditions \eqref{sect5:NBCpsionQ} and \eqref{sect5:NBCpsionE} with $b(0)=0$, we derive
\begin{eqnarray}\label{sect5:orthogonal relationship4}
&&\int_0^{\rho}  b(r)^q a(r)^{p+1} \psi_{m,l}'(r) \psi'_{m,l}(r) \mathrm{d}r\nonumber\\
&=& \int_0^{\rho}  b(r)^{q-2} a(r)^{p+1} \psi^2_{m,l}(r) \left( m^2\frac{b(r)^2}{a(r)^2}-l(l+q-1)\right) \mathrm{d}r\ge 0.
\end{eqnarray}
We have carefully chosen the indexes of $a(r)$ and $b(r)$ for \eqref{sect5:orthogonal relationship3} so that there is no $\psi'_{m,l}(r)$ in Eq. \eqref{sect5:orthogonal relationship4}.

For the $s$-wave with $l=0$, one can easily derive $m^2\ge 0$ from Eq. (\ref{sect5:orthogonal relationship4}). The case with non-zero $l$ is somewhat complicated. We notice that we always have $b(r) < a(r)$. For the tensionless brane $E$, we have $b(r)=\sinh(r), a(r)=\cosh(r)$, which obeys $b(r) < a(r)$. As for the tensive case, only the brane $E$ with codimensions two is consistent with Einstein gravity \cite{Bostock:2003cv}. For the codim-2 brane $E$, note that
\begin{align}
  b(r)^2=f(\bar{r})=\bar{r}^2-1-\left(\frac{\bar{r}_h}{\bar{r}}\right)^{d-2}\left(\bar{r}_h^2-1\right)\,, \quad a(r)^2=\bar{r}^2\,,
\end{align}  
which also satisfies $b(r) < a(r)$. By applying $b(r) < a(r)$, we can prove 
\begin{eqnarray}\label{sect5: positive mass}
m^2\ge l(l+q-1) \ge 0.
\end{eqnarray}
Supposing $m^2<l(l+q-1)$, we derive from (\ref{sect5:orthogonal relationship4}) a contradiction
\begin{eqnarray}\label{sect5: positive mass1}
&&\int_0^{\rho} b(r)^q a(r)^{p+1} \psi_{m,l}'(r) \psi'_{m,l}(r) \mathrm{d}r \ge 0 \nonumber\\
&<& l(l+q-1) \int_0^{\rho} b(r)^{q-2} a(r)^{p+1} \psi^2_{m,l}(r) \left(\frac{b(r)^2}{a(r)^2}-1\right) \mathrm{d}r< 0.
\end{eqnarray}
As a result, the inequality (\ref{sect5: positive mass}) must be held. Now, we have proved that the mass spectrum of KK gravitons is non-negative for the general cases. 

Some comments can be summarised as follows.
\begin{itemize}
  \item The massless KK mode appears only in the $s$-wave with $l=0$. From (\ref{sect5: positive mass}), it is clear that $m^2\ge l(l+q-1)>0$ for $l>0$.
  \item The orthogonal relationship (\ref{sect5:orthogonal relationship}) of KK gravitons is not priori. It holds only for suitable boundary conditions. Fortunately, the $p$-form fields do not affect the boundary conditions (\ref{sect5:NBCpsionQ}) and thus the orthogonal relationship (\ref{sect5:orthogonal relationship}). On the other hand, the DGP gravity and brane-localized higher derivative gravity modify the boundary conditions and the corresponding orthogonal relationship \cite{Miao:2023mui}. As a result, $m^2$ can be negative and even complex for the negative DGP gravity and some kinds of brane-localized higher derivative gravity \cite{Miao:2023mui}.
  \item The real and non-negative mass spectrum of KK gravitons suggests that cone holography with NBC/MBC is stable under linear tensor perturbations. In general, $m^2$ can be negative in AdS$_{p+1}$ as long as it obeys the Breitenlohner-Freedman (BF) bound $m^2\ge -\left(\frac{p}{2}\right)^2$~\cite{Breitenlohner:1982bm}. Since the mass spectrum of cone holography is non-negative, it behaves better than the BF bound.
  \item The massless KK mode is normalized because the coordinate $r$ ranges from $0$ to the position of the EOW brane, and it is localized on the brane $E$ and $Q$. There are a series of massive KK modes of graviton which contribute to the gravity on branes some corrections.
\end{itemize}

\section{Holographic $c$-theorem}
\label{sec:ctheorem}

In this section, we investigate the holographic $c$-theorem in cone holography. Note that the holographic $c$-theorem in cone holography corresponds to the holographic $g$-theorem in AdS/BCFT \cite{Takayanagi:2011zk}. 

To start, we first demonstrate qualitatively why $c$-theorem applies in cone holography. The A-type central charge of $\text{CFT}_p$ is inversely proportional to the effective gravitational constant on the AdS$_{p+1}$ brane \cite{Miao:2021ual}
\begin{align}\label{sect6: a charge}
c \sim \frac{1}{G_{\text{eff}}}=\frac{1}{G_{\text{N}}}\int_{0}^{\rho} b(r)^{q} a(r)^{p-1} \mathrm{d}r, 
\end{align}
which increases with $\rho$. Recall that $r=\rho$ denotes the location of EOW brane $Q$; large $\rho$ corresponds to ultra-violet (UV, AdS boundary at $r=\infty$); small $\rho$ corresponds to infra-red (IR, deep bulk). Thus, we have the $c$-theorem
\begin{align}\label{sect6: c theorem first glance }
c_{\text{UV}} \ge c_{\text{IR}}. 
\end{align}

This section proves the holographic $c$-theorem by using NEC in bulk $C$. We follow the approach of \cite{Myers:2010tj,Kobayashi:2018lil}. 

Recall the solution to vacuum Einstein gravity is given by 
\begin{align} \label{sect6: metric UV}
  \mathrm{d}s^2 =\mathrm{d}r^2 +b(r)^2\mathrm{d}\Omega_q^2 +a(r)^2\mathrm{d}s_{\text{AdS}_{p+1}}^2.
\end{align}
Inspired by \cite{Myers:2010tj,Kobayashi:2018lil}, we assume the solution to Einstein gravity coupled with bulk matter fields takes the form
\begin{align} \label{sect6: metric general}
  \mathrm{d}s^2 =\mathrm{d}r^2 +\frac{b(r)^2}{R(z)}\mathrm{d}\Omega_q^2 +\frac{a(r)^2}{R(z)}\mathrm{d}s_{\text{AdS}_{p+1}}^2\,
\end{align}
where $z$ is the coordinate of AdS$_{p+1}$ and labels the renormalization scale. At the UV fixed point $z\to 0$, we require 
\begin{align}
  \lim_{z\to 0}R(z)=1. 
\end{align}
There is a natural speculation of the $c$-function 
\begin{align} \label{sect6: c function in bulk}
  c(z)=&\frac{1}{G_{\text{N}}}\int^{\rho}_0 \left( \frac{b(r)^2}{R(z)} \right)^{\frac{q}{2}} \left( \frac{a(r)^2}{R(z)} \right)^{\frac{p-1}{2}}\mathrm{d}r \nonumber \\
  =&R(z)^{\frac{2-d}{2}}\frac{1}{G_{\text{N}}}\int^{\rho}_0 b(r)^{q} a(r)^{p-1}\mathrm{d}r\,,
\end{align}
which reduces to the central charge \eqref{sect6: a charge} (up to some positive constants) at the UV fixed point $z\to 0$, $R(z) \to 1$. Above we have used $d=q+p+1$. 

Following the standard approach \cite{Myers:2010tj,Kobayashi:2018lil}, we impose NEC for matter fields in bulk $C$
\begin{align} \label{sect6: NEC in bulk}
  T_{AB}(N_{C})^{A}(N_{C})^{B}\geq 0\,,
\end{align}
where $N_{C}$ is the bulk null vector with the non-zero components $(N_{C})^{z}=(N_{C})^{t}=1$. Combing the NEC \eqref{sect6: NEC in bulk} with Einstein equation $G_{AB}=8\pi G_{\text{N}} T_{AB}$, we get a restriction
\begin{align}
  \frac{d-2}{2z^2 \sqrt{R(z)}}\left(\frac{z^2 R'(z)}{\sqrt{R(z)}}\right)'\geq 0.
\end{align}
Since
\begin{align}
  \lim_{z\to 0} \left(\frac{z^2 R'(z)}{\sqrt{R(z)}}\right)=0\,,
\end{align}
the above inequality implies $\left(\frac{z^2 R'(z)}{\sqrt{R(z)}}\right)\ge 0$ and thus
\begin{align}
  R'(z)\geq 0\,.
\end{align}
Taking the derivative of \eqref{sect6: c function in bulk} regarding $z$, we find the $c$-function decrease along the RG flow
\begin{align}\label{dG}
  c'(z)=-\frac{d-2}{2}R(z)^{-\frac{d}{2}}R'(z)\frac{1}{G_{\text{N}}}\int^{\rho}_0 b(r) ^{q} a(r)^{p-1}\mathrm{d}r\leq 0\,.
\end{align}
As a result, we have $c_{\text{UV}} \ge c_{\text{IR}}$ and prove the holographic $c$-theorem. 

To summarize, we have proved the $c$-theorem of cone holography by applying the NEC in bulk. It is a solid support for the cone holography.

\section{Cone holography with NBC II}
\label{sec:coneNBC2}

In this section, we take another approach to formulate cone holography with NBC. For simplicity, we focus on the most interesting case with a codim-$2$ brane $E$ \footnote{ Einstein gravity can not couple with tensive thin branes with codimensions higher than two.}. Our model comprises Maxwell's fields in bulk and massive vector (Proca) fields on the EOW brane. It can also include Chern-Simons terms in bulk, not affecting our discussions. This holographic model is initially developed to derive chiral current near a boundary \cite{Miao:2022oas,Zhao:2023rno}. Interestingly, it can also accomplish NBC for cone holography.

Our model is given by the gravitational action \eqref{eq:action} with $L_Q=0$ plus the following vector action
\begin{align}\label{eq:actiong}
 I_A=\int_C\mathrm{d}^{d+1}x\sqrt{-g}\left(-\frac{1}{4}\mathcal{F}_{AB}\mathcal{F}^{AB}\right)+\int_Q\mathrm{d}^dx\sqrt{-h}\left(-\frac{1}{2}m_A^2h^{\alpha\beta}A_\alpha A_\beta\right).
\end{align}
Here, $\mathcal{A}$ and $A$ are the vector in bulk and induced vector on the EOW brane, $\mathcal{F}=\mathrm{d}\mathcal{A}$ and $F=\mathrm{d}A$ are strength tensors in bulk and on brane $Q$, respectively. There can be a Chern-Simons term $L_{\text{CS}}(\mathcal{A})$ in bulk for odd $(d+1)$ generally. We do not consider it since it is irrelevant to our discussions below \footnote{ We focus on the maximally symmetrical solutions in this section. As a result, the effects of Chern-Simons terms vanish.}. We add a mass term for the induced vector on the EOW brane. This mass term is important in producing the expected boundary chiral current \cite{Miao:2022oas} and realizing NBC for cone holography.  One can also add an intrinsic kinetic energy term $-\frac{\lambda}{4}F_{\alpha\beta} F^{\alpha\beta}$ on the EOW brane $Q$ \cite{Zhao:2023rno}. Since we focus on the constant induced vector on $Q$ below, the kinetic energy term is irrelevant. 

From the action \eqref{eq:actiong}, we derive NBC for the vector
\begin{align} \label{axialvectorNBC}
n_{A}\mathcal{F}^{AB}+m_A^2 A^{\beta} h^{B}_{\beta} =0,
\end{align} 
where $n_A$ is the outpointing unit normal vector on EOW brane, $h^{B}_{\beta}=\frac{\partial X^B}{\partial x^{\beta}}, A_{\beta}= h^{B}_{\beta}\mathcal{A}_B, A^{\beta}=h^{\beta\alpha}A_{\alpha}$. The energy-momentum tensor for the massive vector field on the EOW brane $Q$ reads
\begin{align}\label{eq:emt-massivevector}
  T_{\alpha\beta}=&\frac{1}{2}m_A^2\left(A_\alpha A_\beta-\frac{1}{2}h_{\alpha\beta}A_\lambda A^\lambda\right)\,.
\end{align}

There is an analytical solution to the Einstein-Maxwell equations in bulk, which reads
\begin{align}\label{eq:calA}
 \mathcal{A}_B=\left(0,\sqrt{\frac{2(d-1)}{d-2}}\frac{Q_{\text{e}}}{\bar{r}^{d-2}},0,\cdots,0\right)\,,
\end{align}
and 
\begin{align}\label{eq:linee4}
 \mathrm{d}s^2=\frac{1}{f(\bar{r})}\mathrm{d}\bar{r}^2+f(\bar{r})\mathrm{d}\theta^2+\bar{r}^2\gamma_{ij}\mathrm{d}x^i\mathrm{d}x^j\,,
\end{align}
where $Q_{\text{e}}$ is the ``electric charge'', $\gamma_{ij}$ is the AdS metric with a unit radius, and the blacking factor is given by
\begin{align}\label{eq:frbarQ}
 f(\bar{r})=\bar{r}^2-1-\frac{Q_{\text{e}}^2}{\bar{r}^{2(d-2)}}-\left(\frac{\bar{r}_h}{\bar{r}}\right)^{d-2}\left(\bar{r}_h^2-1-\frac{Q_{\text{e}}^2}{\bar{r}_h^{2(d-2)}}\right)\,, \quad \bar{r}_h\leq \bar{r}\leq \bar{\rho}\,.
\end{align}
Note that we have $f(\bar{r}_h)=0$, and the codimension-$2$ brane $E$ and EOW brane $Q$ locate at $\bar{r}=\bar{r}_h$ and $\bar{r}=\bar{\rho}$, respectively. The solution \eqref{eq:linee4} resembles the charged hyperbolic black hole \cite{Belin:2013uta}. The difference between them is that the time coordinate of the black hole is replaced by the coordinate $\theta$. As a result, the ``electric charge'' $Q_{\text{e}}$ of \cite{Belin:2013uta} is replaced by $\mathrm{i} Q_{\text{e}}$ in  (\ref{eq:frbarQ}). Strictly speaking, $Q_{\text{e}}$ of (\ref{eq:calA}) is the electric current rather than charge. 

From the metric \eqref{eq:linee4}, we derive the components of extrinsic curvature
\begin{align}\label{eq:extrinsic_curvature1}
  K_{\theta\theta} =& \frac{1}{2}\frac{\partial h_{\theta\theta}}{\partial r}|_Q= \frac{\sqrt{f(\bar{\rho})}}{2}\frac{\partial h_{\theta\theta}}{\partial \bar{\rho}}=\frac{\sqrt{f(\bar{\rho})}}{2} f'(\bar{\rho})=\frac{1}{2}\frac{f'(\bar{\rho})}{\sqrt{f(\bar{\rho})}}h_{\theta\theta}\,, \\
  K_{ij} =& \frac{1}{2}\frac{\partial h_{ij}}{\partial r}|_Q= \frac{\sqrt{f(\bar{\rho})}}{2}\frac{\partial h_{ij}}{\partial \bar{\rho}}=\sqrt{f(\bar{\rho})}\bar{\rho}\gamma_{ij} =\frac{\sqrt{f(\bar{\rho})}}{\bar{\rho}}h_{ij}\,, \label{eq:extrinsic_curvature2}
\end{align}
where we have used $\mathrm{d}r=\mathrm{d}\bar{r}/\sqrt{f(\bar{r})}$. The trace of the extrinsic curvature reads
\begin{align}\label{eq:ec-scalar}
  K =& h^{\theta\theta}K_{\theta\theta}+h^{ij}K_{ij} = \frac{1}{2}\frac{f'(\bar{\rho})}{\sqrt{f(\bar{\rho})}}+(d-1)\frac{\sqrt{f(\bar{\rho})}}{\bar{\rho}} \,.
\end{align}

From Eq. \eqref{eq:calA}, we read off the induced vector on the EOW brane $\bar{r}=\bar{\rho}$, 
\begin{align}\label{eq:massivevector}
  A_{\alpha}=\left(A_\theta,0,\cdots,0\right)=\left(\sqrt{\frac{2(d-1)}{d-2}}\frac{Q_{\text{e}}}{\bar{\rho}^{d-2}},0,\cdots,0\right)\,.
\end{align}
It shows that $A_{\alpha}$ is a constant vector with only non-zero component $A_{\theta}$. Then the energy-momentum tensor \eqref{eq:emt-massivevector} becomes
\begin{align}\label{eq:emt-massivevectorabij}
  T_{\theta\theta}=\frac{1}{4}m_A^2\left(A_{\theta} A_{\theta} h^{\theta\theta}\right) h_{\theta\theta}\,, \quad  T_{ij}=-\frac{1}{4}m_A^2\left(A_{\theta} A_{\theta} h^{\theta\theta}\right) h_{ij},
\end{align}
where $\left(A_{\theta} A_{\theta} h^{\theta\theta}\right) =A^2_{\theta}/f(\bar{\rho}) $ is a constant. We remark that $T_{\theta\theta}$ and $T_{ij}$ are taken the expected asymmetry to compensate that of extrinsic curvatures \eqref{eq:extrinsic_curvature1} and \eqref{eq:extrinsic_curvature2} and to obey the  NBC~\eqref{eq:NBC}.

Substituting the extrinsic curvatures \eqref{eq:extrinsic_curvature1}, \eqref{eq:extrinsic_curvature2} and vector energy-momentum tensor \eqref{eq:emt-massivevectorabij} into NBC~\eqref{eq:NBC}, we get two independent equations from the $\left(\theta,\theta\right)$ and $\left(i,j\right)$ components. Recall that we have two free parameters $T_Q$ and $m_A^2$, which happens to solve the two independent equations. We obtain
\begin{align}\label{eq:solution4a}
T_Q=&\frac{(4 d-6) f\left(\bar{\rho }\right)+\bar{\rho } f'\left(\bar{\rho }\right)}{4 \bar{\rho } \sqrt{f\left(\bar{\rho }\right)}},\\
m_A^2=&\frac{\sqrt{f\left(\bar{\rho }\right)} \left(\bar{\rho } f'\left(\bar{\rho }\right)-2 f\left(\bar{\rho }\right)\right)}{\bar{\rho } A^2_{\theta }},\label{eq:solution4b}
\end{align}
where $A_{\theta}=\sqrt{\frac{2(d-1)}{d-2}}\frac{Q_{\text{e}}}{\bar{\rho}^{d-2}}$ is a constant. Above we focus on the NBC~\eqref{eq:NBC} with respect to gravity. The NBC~(\ref{axialvectorNBC}) of vectors set another constraint 
\begin{align}\label{sect7:massfromvectorNBC}
m_A^2=\frac{(d-2)\sqrt{f\left(\bar{\rho }\right)} }{\bar{\rho }}.
\end{align}
Comparing (\ref{eq:solution4b}) with (\ref{sect7:massfromvectorNBC}), we observe that the charge $Q_{\text{e}}$ and brane location $\bar{\rho}$ are not independent, which is similar to the case of $p$-form fields. We obtain 
\begin{align}\label{sect7:charge}
Q_{\text{e}}^2=\bar{r}_h^{2 d-4} \left(\frac{2}{d} \left(\frac{\bar{\rho }}{\bar{r}_h}\right){}^{d-2}+\bar{r}_h^2-1\right).
\end{align}
Now we have shown that our model with massive vectors on the EOW brane can indeed enable NBC.

Some comments are listed as follows. First, the mass squared (\ref{sect7:massfromvectorNBC}) is positive $m_A^2>0$, which implies this model is stable and tachyon-free. Second, there is no well-defined charge-free limit $Q_{\text{e}}\to 0$. The reason is as follows. From (\ref{eq:frbarQ}) with $Q_{\text{e}}=0$, we derive the reciprocal of angle period 
\begin{align}\label{sect7:angle period}
\lim_{Q_{\text{e}}\to 0}\frac{1}{2\pi n}=\frac{f'(\bar{r}_h)}{4\pi}=\frac{d \left(\bar{r}_h^2-1\right)+2}{4 \pi  \bar{r}_h}\ge 0,
\end{align}
which yields $\sqrt{\frac{d-2}{d}}\le \bar{r}_h$. On the other hand, we obtain from (\ref{sect7:charge})
\begin{align}\label{sect7:zero charge}
\lim_{Q_{\text{e}}\to 0}\bar{\rho}=\bar{r}_h\left(\frac{1}{2} \left(d-d \bar{r}_h^2\right)\right){}^{\frac{1}{d-2}} \le \bar{r}_h,
\end{align}
for $\sqrt{\frac{d-2}{d}}\le \bar{r}_h$. It leads to a contradiction, since  the EOW brane $Q$ should lie outside the brane $E$, i.e, $\bar{\rho} > \bar{r}_h$. It suggests the charge is necessary for cone holography with NBC. Interestingly, this is also the case for models with brane-localized $p$-form fields in sections~\ref{sec:toymodel} and \ref{sec:coneNBC1}.

\section{Conclusions and discussions}
\label{sec:conclusions}

This paper formulates cone holography with NBC by employing $p$-form fields on the EOW brane. NBC is closely related to the junction condition of branes and a massless graviton on branes is allowed. Most doubly holographic models are based on NBC. Thus, there is a good motivation to construct cone holography with NBC. We observe that the central obstruction for NBC is the asymmetry of the two sectors $S^q$ and AdS$_{p+1}$ of EOW brane $Q$. Negative DGP gravity can resolve this problem but suffers the ghost issue. Remarkably, the brane-localized $p$-form fields can compensate for the asymmetry of the EOW brane and accomplish NBC. Moreover, the $p$-form field takes the standard form and is ghost-free. We have analyzed the perturbation solution and proved that the mass spectrum is non-negative. We also have proven the holographic $c$-theorem for cone holography with NBC. These are solid supports for our models. Finally, inspired by the chiral theory in AdS/BCFT, we have constructed another model of cone holography with NBC by applying a massive vector (Proca) field on the EOW brane. The mass of Proca field is positive, implying that this model is stable. Interestingly, the charge is necessary to enable NBC in both models.  

Let us make two further comments. First, the $p$-form fields appear naturally in string theory. It raises the question of whether cone holography has a string theory origin. We leave this interesting problem to future work. Second, it is interesting to calculate the generalized gravitational entropy (holographic entanglement entropy) of cone holography with brane-localized $p$-form fields. The generalized gravitational entropy is given by the Ryu-Takayanagi (RT) formula \cite{Ryu:2006bv} in bulk, without corrections from the brane-localized $p$-form fields. That is because the $p$-form fields on the EOW brane is classical since it appears in the classical Neumann boundary condition. Thus, it is unnecessary to consider its quantum entanglement entropy from the bulk perspective. On the other hand, from the brane perspective, we have effective higher derivative gravity plus quantum matter fields on the brane. As a result, the entanglement entropy is given by the generalized RT formula of higher derivative gravity \cite{Dong:2013qoa,Camps:2013zua,Miao:2014nxa} plus von Neumann entropy of quantum matters. Note that the classical $p$-form fields in the bulk description differ from the quantum matters in the brane description. The bulk and brane entropy formulas are expected to be equivalent. But the bulk calculation is more doable. We remark that, in general, the brane-localized $p$-form fields back-react the geometry through the Neumann boundary condition. Thus, although it does not affect the bulk entropy formula, it generally contributes to the entropy.

\section*{Acknowledgements}

We thank Dong-Qi Li for valuable discussions. Rong-Xing Miao is supported by the National Natural Science Foundation of China (No. 12275366 and No. 11905297). Zheng-Quan Cui is supported by the National Natural Science Foundation of China (No. 12247135).

\appendix

\section{Conventions}
\label{app:a}

In this Appendix we list conventions adopted in this paper. We are using the definitions $R^P_{MQN}=\partial_Q\Gamma^P_{MN}-\partial_N\Gamma^P_{MQ}+\Gamma^P_{QL}\Gamma^L_{MN}-\Gamma^P_{NL}\Gamma^L_{MQ}$ and $R_{MN}=R^L_{MLN}$. The metric signature is the mostly plus convention $(-,+,+,\cdots,+)$. We will use the natural units with $16\pi G_{\text{N}}=c=\hbar=1$, where $G_{\text{N}}$ is the gravitational constant. We use the unit normalized convention for symmetrization and antisymmetrization. Parentheses $(\,)$ which appear in indices denote symmetrization, for example
\[A_{(\mu} B_{\nu)}=\frac{1}{2}(A_\mu B_\nu+A_\nu B_\mu),\quad A_{(\mu\nu)}=\frac{1}{2}(A_{\mu\nu}+A_{\nu\mu}).\]
Square brackets $[\,]$ which appear in indices denote antisymmetrization, for example
\[A_{[\mu} B_{\nu]}=\frac{1}{2}(A_\mu B_\nu-A_\nu B_\mu),\quad A_{[\mu\nu]}=\frac{1}{2}(A_{\mu\nu}-A_{\nu\mu}).\]

\section{Perturbations of spacetime quantities}
\label{app:b}

On the EOW brane $Q$, the linear tensor perturbations which we consider is
\begin{align}\label{eq:lineeg-perturbation}
  \mathrm{d}s^2=\mathrm{d}r^2+b(r)^2\left(\vartheta_{ab}+H_{ab}\right)\mathrm{d}x^ax^b+a(r)^2\left(\gamma_{ij}+H_{ij}\right)\mathrm{d}x^i \mathrm{d}x^j\,.
\end{align}
Here, these tensor perturbations satisfies TT condition
\begin{align}\label{}
  \nabla^{(\vartheta)}_a H^a_b=0\,, \quad H\equiv\vartheta^{ab}H_{ab}=0\,, \quad \nabla^{(\gamma)}_i H^i_j=0\,, \quad H\equiv\gamma^{ij}H_{ij}=0\,. 
\end{align}
We use $\nabla^{(\vartheta)}$ and $\nabla^{(\gamma)}$ to signify the covariant derivative defined by $\vartheta_{ab}$ and $\gamma_{ij}$. Generally, the tensor perturbations $H_{ab}$ and $H_{ij}$ are the functions of all spacetime coordinates $\left(r,x^a,x^i\right)$. Why we can consider such an ansatz is that a general perturbation $H_{\alpha\beta}$ around the background metric $h_{\alpha\beta}$ will destroy the configuration of cone holography and is forbad.

From Eq.~\eqref{eq:general-perturbations}, the spacetime quantities can be separated into
\begin{align}
  \tilde{R}_{ab}= R_{ab}+\delta R_{ab}\,, \quad \tilde{R}_{ij}= R_{ij}+\delta R_{ij}\,, \quad \tilde{R}=R+\delta R\,.
\end{align}
The background components of Ricci tensor and Ricci scalar are
\begin{align}\label{aeq:Riccitensor}
  R_{ab}=&R^{(\vartheta)}_{ab}-bb'\left[(p+1)\frac{a'}{a}+q\frac{b'}{b}\right]\vartheta_{ab}-b^2\partial_r\left(\frac{b'}{b}\right)\vartheta_{ab}\,,\\
  R_{ij}=&R^{(\gamma)}_{ij}-aa'\left[(p+1)\frac{a'}{a}+q\frac{b'}{b}\right]\gamma_{ij}-a^2\partial_r\left(\frac{a'}{a}\right)\gamma_{ij}\,, \\
  R_{rr}=&-(p+1)\frac{a''}{a}-q\frac{b''}{b}\,,\\
  R = &\frac{1}{a^2}R^{(\gamma)}+\frac{1}{b^2}R^{(\vartheta)} -(p+1)\frac{pa'^2+2aa''}{a^2}-q\frac{(q-1)b'^2+2bb''}{b^2} -2(p+1)q\frac{a'}{a}\frac{b'}{b}\,.
\end{align}
Here the prime denotes the derivative with respect to the coordinate $r$, and $R^{(\vartheta)}$ and $R^{(\gamma)}$ are the Ricci scalars constructed by the metric $\vartheta_{ab}$ and $\gamma_{ij}$ respectively.
The perturbations of Ricci tensor and Ricci scalar are calculated as
\begin{align}\label{aeq:Ricci-componnets}
  \delta R_{ab}=& -\frac{1}{2}\frac{b^2}{a^2}\square^{(\gamma)}H_{ab}-\frac{1}{2}\Delta^{(\vartheta)}H_{ab}-\frac{1}{2}b^2\partial_r\partial_rH_{ab}
  -\frac{1}{2}b^2\left[(p+1)\frac{a'}{a}+q\frac{b'}{b}\right]\partial_r H_{ab}\nonumber\\ &+\frac{1}{q-1}R^{(\vartheta)}H_{ab}-bb'\left[(p+1)\frac{a'}{a}+q\frac{b'}{b}\right]H_{ab}-b^2\partial_r\left(\frac{b'}{b}\right)H_{ab}\,,\\
  \delta R_{ij}=& -\frac{1}{2}\square^{(\gamma)}H_{ij}-\frac{1}{2}\frac{a^2}{b^2}\Delta^{(\vartheta)}H_{ij}-\frac{1}{2}a^2\partial_r\partial_rH_{ij}
  -\frac{1}{2}a^2\left[(p+1)\frac{a'}{a}+q\frac{b'}{b}\right]\partial_r H_{ij}\nonumber\\ &+\frac{1}{p}R^{(\gamma)}H_{ij}-aa'\left[(p+1)\frac{a'}{a}+q\frac{b'}{b}\right]H_{ij}-a^2\partial_r\left(\frac{a'}{a}\right)H_{ij}\,,\\
  \delta R=& 0\,.
\end{align}
Here $\square^{(\gamma)}=\gamma^{ij}\nabla^{(\gamma)}_i\nabla^{(\gamma)}_j$ and $\Delta^{(\vartheta)}=\vartheta^{ab}\nabla^{(\vartheta)}_{a}\nabla^{(\vartheta)}_{b}$ denote d'Alembert operators in the AdS$_{p+1}$ and the unit sphere $S^q$, respectively.

From the Einstein equation
\begin{align}\label{aeq:Einsteineq}
  R_{AB} = \frac{2}{d-1}\Lambda g_{AB}\,,  
\end{align}
we have $(a,b)$ and $(i,j)$ components of background Einstein equations
\begin{align}\label{aeq:Einsteineqs-background}
  R_{ab} = \frac{2}{d-1}\Lambda b^2 \vartheta_{ab}\,, \quad   R_{ij} = \frac{2}{d-1}\Lambda a^2 \gamma_{ij}\,,
\end{align}
and its perturbation equations
\begin{align}\label{aeq:Einsteineqs-perturbation}
  \delta R_{ab} = \frac{2}{d-1}\Lambda b^2 H_{ab}\,, \quad  \delta R_{ij} = \frac{2}{d-1}\Lambda a^2 H_{ij}\,.
\end{align}
The concrete background Einstein equations are
\begin{align}\label{aeq:Einsteineqs-backgroundc}
  \frac{1}{b^2} R^{(\vartheta)} -q\frac{(q-1)b'^2+bb''}{b^2}-(p+1)q\frac{a'}{a}\frac{b'}{b}=& q\frac{2}{d-1}\Lambda\,, \\
  \frac{1}{a^2} R^{(\gamma)}-(p+1)\frac{pa'^2+aa''}{a^2}-(p+1)q\frac{a'}{a}\frac{b'}{b} = & (p+1)\frac{2}{d-1}\Lambda\,, \\
  -(p+1)\frac{a''}{a}-q\frac{b''}{b}=& \frac{2}{d-1}\Lambda\,.
\end{align}
The concrete perturbation equations are given by
\begin{align}\label{aeq:Einsteineqs-perturbationc}
  \frac{1}{a^2}\square^{(\gamma)}H_{ab}+ \frac{1}{b^2}\left(\Delta^{(\vartheta)}-\frac{2}{q(q-1)}R^{(\vartheta)}\right)H_{ab}+\partial_r\partial_r H_{ab}+
  \left[(p+1)\frac{a'}{a}+q\frac{b'}{b}\right]\partial_r H_{ab}=0\,, \\
  \frac{1}{a^2}\left(\square^{(\gamma)}-\frac{2}{p(p+1)}R^{(\gamma)}\right)H_{ij}+ \frac{1}{b^2}\Delta^{(\vartheta)}H_{ij}+\partial_r\partial_r H_{ij}+
  \left[(p+1)\frac{a'}{a}+q\frac{b'}{b}\right]\partial_r H_{ij} =0\,. 
\end{align}

\bibliographystyle{jhep}  

\end{document}